\documentclass[numberedappendix]{emulateapj}

\usepackage{graphicx}

\def\kms{\hbox{$~$km$~$s$^{-1}$}}
\def\l{\ifmmode\lambda\else$\lambda$\fi}
\def\snia{SN~Ia}
\def\sneia{SNe~Ia}
\newcommand{\beq}{\begin{equation}}
\newcommand{\eeq}{\end{equation}}
\newcommand{\bea}{\begin{eqnarray}}
\newcommand{\eea}{\end{eqnarray}}

\slugcomment{Accepted for publication in Ap.~J.}
\shorttitle{Time Dilation in SN~Ia Spectra}
\shortauthors{Blondin et al.}

\begin{document}

\title{Time Dilation in Type I\lowercase{a} Supernova Spectra
  at High Redshift\footnotemark[1]}

\footnotetext[1]{\vspace{0.00cm}Based on observations made with ESO
  Telescopes at the Paranal Observatory under programs 67.A-0361,
  267.A-5688, 078.D-0383, and 080.D-0477; at the Gemini Observatory and
  NOAO, which are operated by the Association of Universities for
  Research in Astronomy, Inc., under cooperative agreements with the
  NSF; with the Magellan Telescopes at Las Campanas Observatory; and at the
  W.~M.~Keck Observatory, which was made possible by the generous
  financial support of the W.~M.~Keck Foundation.}

\author{
S.~Blondin,\altaffilmark{2}
T.~M.~Davis,\altaffilmark{3,4}
K.~Krisciunas,\altaffilmark{5} 
B.~P.~Schmidt,\altaffilmark{6}
J.~Sollerman,\altaffilmark{3} 
W.~M.~Wood-Vasey,\altaffilmark{2} 
A.~C.~Becker,\altaffilmark{7} \\
P.~Challis,\altaffilmark{2}
A.~Clocchiatti,\altaffilmark{8}
G.~Damke,\altaffilmark{9} 
A.~V.~Filippenko,\altaffilmark{10}
R.~J.~Foley,\altaffilmark{10} 
P.~M.~Garnavich,\altaffilmark{11} 
S.~W.~Jha,\altaffilmark{12} \\
R.~P.~Kirshner,\altaffilmark{2} 
B.~Leibundgut,\altaffilmark{13}
W.~Li,\altaffilmark{10} 
T.~Matheson,\altaffilmark{14}
G.~Miknaitis,\altaffilmark{15}
G.~Narayan,\altaffilmark{16} 
G.~Pignata,\altaffilmark{17} \\ 
A.~Rest,\altaffilmark{9,16} 
A.~G.~Riess,\altaffilmark{18,19}
J.~M.~Silverman,\altaffilmark{10}
R.~C.~Smith,\altaffilmark{9}  
J.~Spyromilio,\altaffilmark{13} 
M.~Stritzinger,\altaffilmark{3,20} \\
C.~W.~Stubbs,\altaffilmark{2,16}
N.~B.~Suntzeff,\altaffilmark{5}
J.~L.~Tonry,\altaffilmark{21}
B.~E.~Tucker,\altaffilmark{6} 
and A.~Zenteno\altaffilmark{22}
}

\altaffiltext{2}{Harvard-Smithsonian Center for Astrophysics, 60
  Garden Street, Cambridge, MA 02138; {sblondin@cfa.harvard.edu} }
\altaffiltext{3}{Dark Cosmology Centre, Niels Bohr Institute, University of Copenhagen, Juliane Maries Vej 30, DK--2100 Copenhagen~\O, Denmark.} 
\altaffiltext{4}{Department of Physics, University of Queensland, QLD, 4072, Australia.}
\altaffiltext{5}{Department of Physics, Texas A\&M University, College Station, TX 77843-4242.}  
\altaffiltext{6}{The Research School of Astronomy and Astrophysics, The Australian National University, Mount Stromlo and Siding Spring Observatories, via Cotter Road, Weston Creek, PO 2611, Australia.}  
\altaffiltext{7}{Department of Astronomy, University of Washington, Box 351580, Seattle, WA 98195-1580.}  
\altaffiltext{8}{Pontificia Universidad Cat\'olica de Chile, Departamento de Astronom\'ia y Astrof\'isica, Casilla 306, Santiago 22, Chile.}
\altaffiltext{9}{Cerro Tololo Inter-American Observatory, National Optical Astronomy Observatory, Casilla 603, La Serena, Chile.} 
\altaffiltext{10}{Department of Astronomy, University of California, Berkeley, CA 94720-3411.}  
\altaffiltext{11}{Department of Physics, University of Notre Dame, 225 Nieuwland Science Hall, Notre Dame, IN 46556-5670.}  
\altaffiltext{12}{Department of Physics and Astronomy, Rutgers University, 136 Frelinghuysen Road, Piscataway, New Jersey 08854.}
\altaffiltext{13}{European Southern Observatory, Karl-Schwarzschild-Strasse 2, D-85748 Garching, Germany.}
\altaffiltext{14}{National Optical Astronomy Observatory, 950 North Cherry Avenue, Tucson, AZ 85719-4933.}  
\altaffiltext{15}{Fermilab, P.O. Box 500, Batavia, IL 60510-0500.} 
\altaffiltext{16}{Department of Physics, Harvard University, 17 Oxford Street, Cambridge, MA 02138.}
\altaffiltext{17}{Departamento de Astronom{\'{\i}}a, Universidad de Chile, Casilla 36-D, Santiago, Chile.}
\altaffiltext{18}{Johns Hopkins University, 3400 North Charles Street, Baltimore, MD 21218.}  
\altaffiltext{19}{Space Telescope Science Institute, 3700 San Martin Drive, Baltimore, MD 21218.} 
\altaffiltext{20}{Las Campanas Observatory, Carnegie Observatories, Casilla 601, La Serena, Chile.} 
\altaffiltext{21}{Institute for Astronomy, University of Hawaii, 2680 Woodlawn Drive, Honolulu, HI 96822.}  
\altaffiltext{22}{Department of Astronomy, University of Illinois at Urbana-Champaign, 1002 West Green St, Urbana, IL 61801-3080.}


\begin{abstract}
We present multiepoch spectra of 13 high-redshift Type Ia supernovae
(\sneia) drawn from the literature, the ESSENCE and SNLS projects, and
our own separate dedicated program on the ESO Very Large Telescope. We
use the Supernova Identification (SNID) code of Blondin \& Tonry to
determine the spectral ages in the supernova rest frame. Comparison
with the observed elapsed time yields an apparent aging rate
consistent with the $1/(1+z)$ factor (where $z$ is the redshift)
expected in a homogeneous, isotropic, expanding universe. These
measurements thus confirm the expansion hypothesis, while
unambiguously excluding models that predict no time dilation, such as
Zwicky's ``tired light'' hypothesis. 
We also test for power-law dependencies of the aging rate on
redshift. The best-fit exponent for these models is consistent with the
expected $1/(1+z)$ factor.
\end{abstract}



\section{Introduction}\label{sect:intro}

The redshift, $z$, is a fundamental observational quantity in
Friedman-Lema{\^{\i}}tre-Robertson-Walker (FLRW) models of the
universe. It relates the frequency of light emitted from a distant
source to that detected by a local observer by a factor $1/(1+z)$. One
important consequence is that the observed 
rate of any time variation in the intensity of emitted radiation will
also be proportional to $1/(1+z)$ (see \citealt{Weinberg:1972} and
Appendix~\ref{sect:math}). 

Due to their large luminosities (several billion times that of the sun)
and variability on short timescales ($\sim20$~d from explosion
  to peak luminosity; \citealt{Riess/etal:1999c,Conley/etal:2006b}), Type
Ia supernovae 
(\sneia) are ideally suited to probe these time-dilation effects
across a large fraction of the observable universe. The suggestion to
use supernovae as ``cosmic clocks'' was proposed by Wilson more than six
decades ago \citep{Wilson:1939} and tested on light curves of
low-redshift \sneia\ in the mid-1970s \citep{Rust:1974}, but only
since the mid-1990s has this effect been unambiguously detected in the light
curves of high-redshift objects \citep{SN1995K,Goldhaber/etal:2001}.

\begin{figure}
\epsscale{1.1}
\plotone{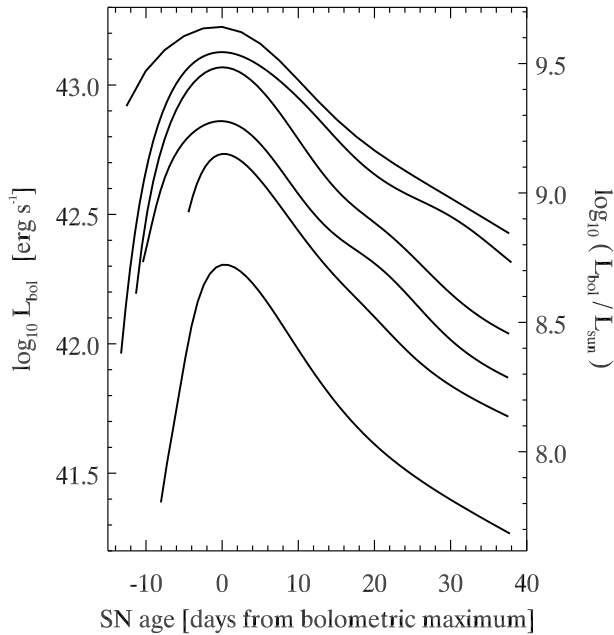}
\caption{\label{fig:lcbol}
Bolometric light
curves of 5 low-redshift \sneia\ taken from \cite{Stritzinger:2005}
(from top to bottom: SNe~1991T, 1999ee, 1994D, 1992A, 1993H, and
1991bg). More luminous \sneia\ have broader light curves. 
SN~1991bg is an example of intrinsically subluminous \sneia\
  (maximum $L_{\rm bol} < 10^9 L_{\rm sun}$), which are less likely to
be found at high redshifts.}
\end{figure}

These latter studies show that the light curves of distant \sneia\ are
consistent with those of nearby \sneia\ whose time axis is dilated by
a factor $(1+z)$. However, there exists an intrinsic variation in the
width of \snia\ light curves which is related to their peak
luminosities \citep{Phillips:1993}, such that more luminous \sneia\
have broader light curves (Fig.~\ref{fig:lcbol}). This
``width-luminosity'' relation is derived using low-redshift \sneia\
where the time-dilation effect, if any, is negligible 
\citep{Phillips:1993,Hamuy/etal:1995,Riess/Press/Kirshner:1995,Phillips/etal:1999,MLCS2k2}.

It is problematic to disentangle this intrinsic variation of
light-curve width with luminosity and the effect of time dilation.
To directly test the time-dilation hypothesis one needs to accurately
know the distribution of light-curve widths at $z\approx0$ and its
potential evolution with redshift, whether due to a selection effect
(not taken into account by \citealt{Goldhaber/etal:2001}) or an evolution
of the mean properties of the \snia\ sample with redshift--- as
possibly observed by \cite{Howell/etal:2007}. Moreover, one needs to
probe sufficiently high redshifts ($z\gtrsim0.4$, as done by
\citealt{SN1995K,Goldhaber/etal:2001}) such that the observed widths of
the \snia\ light curves are broader than the intrinsic width of any
nearby counterpart.

Furthermore, one might argue that
at high redshift we are preferentially finding the
brighter events (akin to a Malmquist bias).
Such a selection effect would produce a spurious relation in
which there would be broader light curves at higher redshifts,
without any time dilation.

The spectra of \sneia\ provide an alternative and a more
reliable way to measure the apparent aging rate of distant
objects. Indeed, the spectra of \sneia\ are remarkably homogeneous at a
given age, such that the age of a \snia\ can be determined from a
single spectrum with an accuracy of 1--3 d--- with no reference to
its corresponding light curve \citep{Riess/etal:1997a,Howell/etal:2005,SNID}.
More importantly, the spectra of \sneia\ spanning a range of luminosities
(and hence different intrinsic light-curve widths) evolve uniformly
over time \citep{Matheson/etal:2008}.
The use of spectra thus avoids
the degeneracy between intrinsic light-curve width and time-dilation
effects. While there are some notable examples of deviations from
homogeneity in several \sneia\ (e.g., SN~2000cx, \citealt{Li/etal:2001b};
SN~2002cx, \citealt{Li/etal:2003}; SN~2002ic, \citealt{Hamuy/etal:2003};
SN~2003fg, \citealt{03D3bb}; SN~2006gz, \citealt{SN2006gz}), these
  outliers are readily identifiable spectroscopically through
comparison with a large database of supernova spectra (see
Section~\ref{sect:age} and \citealt{SNID}).

As of today there are two published examples of aging rate
measurements using spectra of a single \snia\ (SN~1996bj
at $z=0.574$, \citealt{Riess/etal:1997a}; SN~1997ex at $z=0.362$,
\citealt{Foley/etal:2005}). In both cases, the null hypothesis of no time
dilation is excluded with high significance ($>95\%$). 

In this paper we present data on 13 high-redshift ($0.28 \le z \le
0.62$) \sneia\ for which we have multiepoch spectra. We
use the Supernova Identification (SNID) code of \citet{SNID} to infer
the age of each spectrum in the supernova rest frame, from which we
determine the apparent aging rate of each \snia. These aging rate
measurements are then used to test the $1/(1+z)$ time-dilation
hypothesis expected in an expanding universe. The data enable us for
the first time to directly test the time-dilation hypothesis {\it over
  a large redshift range}.

This paper is organized as follows. 
In Section~\ref{sect:age} we explain how one can determine the age
of a \snia\ based on a single spectrum and present the SNID algorithm
used for this purpose. The aging rate measurements are presented in
Section~\ref{sect:aging}, and the time-dilation hypothesis (amongst
others) is tested against the data in Section~\ref{sect:tdil}, with
the help of model selection statistics (information
criteria). Conclusions follow in Section~\ref{sect:ccl}.


\section{Determining the Age of a SN~I\lowercase{a} Spectrum}\label{sect:age}

\subsection{SN~I\lowercase{a} Spectral Evolution}

The spectra of \sneia\ consist of blended spectral
lines, with a profile shape characteristic of stellar
outflows. This line profile (also known as a ``P Cygni'' profile)
consists of an emission component symmetric about the line center, and
an absorption component that is blueshifted by the $\sim10,000$\,\kms\
expansion velocity of the SN ejecta \citep{Pinto/Eastman:2000b}. The
expansion also causes a Doppler broadening of both components, such
that a typical spectroscopic feature in \snia\ spectra has a width of
$\sim100$\,\AA.
As the ejecta expand, the photosphere recedes in the comoving frame
of the supernova, such that the spectra probe deeper layers 
of the ejecta with time. Given the homologous nature of the
expansion (velocity proportional to radius), and the chemical
stratification in the SN ejecta
\citep{W7,Stehle/etal:2005,Mazzali/etal:2008}, deeper layers
correspond to lower expansion velocities and an increased 
abundance of iron-peak elements. The impact on the spectra is
twofold. First, the blueshift of \snia\ spectral lines decreases with
time (by as much as $\sim1000$\,\kms\ per day;
\citealt{Benetti/etal:2005,Blondin/etal:2006}). Second, due to the
varying chemical composition at the photosphere, the relative
shapes and strengths of spectral features change on a timescale of
days.

This complex spectral evolution is nonetheless predictable to a large
extent. At a given age, the spectra are remarkably homogeneous among
different ``normal'' \sneia. According to \cite{Li/etal:2001a}, these
constitute $\sim65$\% of the local \snia\ sample, the rest consisting of 
intrinsically subluminous ($\sim15$\%) or overluminous ($\sim20\%$)
events, whose spectra show deviations from those of normal \sneia. Subluminous
\sneia\ are less likely to be found at high redshifts; in fact, no
such object has been spectroscopically confirmed in any high-redshift
supernova search to this day
\citep[e.g.,][]{Matheson/etal:2005,Howell/etal:2005}. In what follows
we only consider normal and overluminous \sneia. In a separate paper
we will show that none of the \sneia\ in the high-redshift sample
presented here (see Section~\ref{sect:tdil}) has a spectrum or light
curve consistent with the subluminous variety of \sneia.

\begin{figure}
\epsscale{1.1}
\plotone{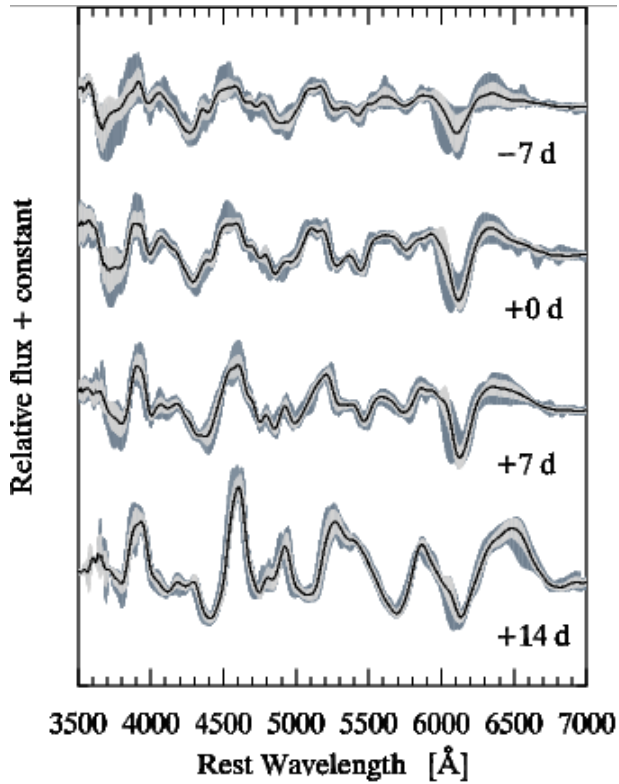}
\caption{\label{fig:meanspec}
Standard (light gray) and maximum (dark gray) deviation from the mean
spectrum (black line) for the 22 low-redshift \sneia\ for which we
report an aging rate measurement (see Section~\ref{sect:tdil}), at
four different ages --- given in days from $B$-band maximum light. A
low-order curve has been divided out from each spectrum to reveal the
relative shapes and strengths of the various spectroscopic features. 
}
\end{figure}

The spectroscopic homogeneity of \sneia\ holds even when we consider
both normal and overluminous objects in a representative sample of
nearby events. In Fig.~\ref{fig:meanspec} we show the mean spectrum
for the 22 low-redshift \sneia\ for which we report an aging rate
measurement (see Section~\ref{sect:tdil}), at four different
ages. While there is an intrinsic spectral variance among these
different \sneia\ --- some spectroscopic features correlate with
luminosity (e.g., \citealt{Nugent/etal:1995,Matheson/etal:2008}), the
average deviations from the mean spectrum are small, and all spectra
evolve in a similar manner over the course of several days, {\it
  independent of light-curve width}.

Both the homogeneity and rapid evolution of \snia\ spectra enables an
accurate determination of the age of a single spectrum. We explain how
this is achieved in practice in the following section.

\subsection{\label{sect:snid}The SNID Algorithm}

Given a large database of finely time-sampled \snia\ spectral
templates, we can determine the age of a given input spectrum by
finding the best-match template(s) in the database. There are several
standard techniques to do this (see \citealt{SNID} for a
review). In this paper, we use an implementation of the correlation
techniques of \cite{Tonry/Davis:1979}: SNID \citep{SNID}.
SNID automatically determines
the type, redshift, and age of a supernova spectrum. We refer the
reader to that paper for a more detailed discussion.

The redshift of the input spectrum is a free parameter in SNID,
although it can be fixed to a specific value. 
Comparison of the SNID redshifts with those determined from
narrow emission and absorption lines in the host-galaxy
spectra (typically accurate to $\lesssim100$\,\kms; see \citealt{UZC})
yields a dispersion about the one-to-one correspondence of only
$\sigma_z \approx 0.005$ out to a redshift $z \approx 0.8$ \citep{SNID}.

Similarly, comparison of the SNID ages with those determined using the
corresponding light curves yields a typical accuracy $<3$ d,
comparable to other algorithms \citep{Riess/etal:1997a,Howell/etal:2005}.
However, the age error is systematically overestimated. In this
paper, we estimate the error as follows: each spectrum in the SNID
database is trimmed to match the rest-frame wavelength range of the
input spectrum, and is correlated with all other spectra in the
database (except those corresponding to the same supernova). The age
error is then given by the mean variance of all template spectra whose
SNID age is within one day of the initial estimate.

\begin{figure}
\epsscale{1.1}
\plotone{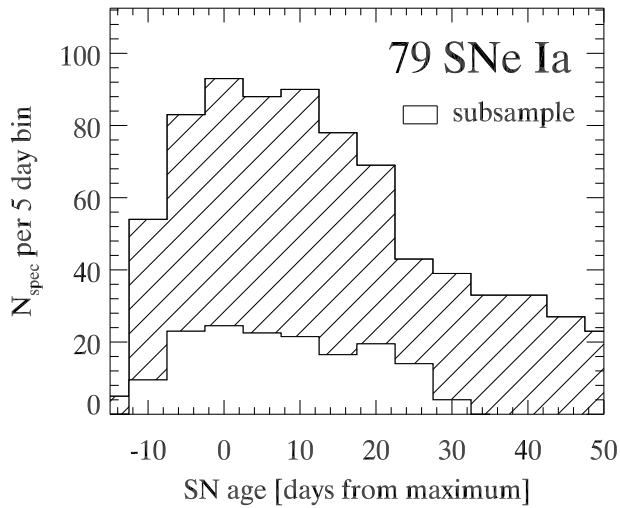}
\caption{\label{fig:thist}
Age distribution of the 79 \snia\ templates used in SNID (hatched
histogram). There are a total of 959 spectra with ages between $-15$
and $+50$~d from $B$-band maximum light. The open histogram shows
the subsample of 145 spectra from 22 \sneia\ (restricted to ages
between $-10$ and $30$~d from maximum) for which we report an aging
rate measurement (see Section~\ref{sect:tdil}).}
\end{figure}

The success of SNID and similar algorithms lies primarily in the
completeness of the spectral database. In Fig.~\ref{fig:thist} we
show the age distribution of the \snia\ templates used in SNID for
this paper (these do not include subluminous \sneia). This database
comprises 959 spectra of 79 low-redshift ($z \lesssim 0.05$) \sneia\
with ages between $-15$ and $+50$~d from maximum light. The spectra
are taken from the literature, from public databases (such as
SUSPECT\footnote{http://bruford.nhn.ou.edu/$\sim$suspect/index1.html}
or the CfA Supernova
Archive\footnote{http://www.cfa.harvard.edu/supernova/SNarchive.html}),
or from a set of unpublished spectra from the CfA Supernova Program. A
full reference to all spectra in the SNID database is given by
\cite{SNID}. It is important to note that each template
spectrum is shifted to zero redshift and that each template age is
corrected for the expected $(1+z)$ time-dilation factor. Because all
the template \sneia\ are at low redshift ($z \lesssim 0.05$), this is
a very small correction, and we will see in Section~\ref{sect:tdil}
that this has no impact on the aging rate measurements. 
Thus, SNID determines the age a supernova would have at $z=0$ --- that is,
{\it in the supernova rest frame}.

The number of \sneia\ shown in Fig.~\ref{fig:thist} is 
large enough that we can select a subsample (shown as an
open histogram) on which to conduct age determinations and aging rate
measurements on low-redshift \sneia. The size of this subsample
is set by the requirement that removing it from the SNID database
would leave a sufficient number of templates in a given age bin for a
reliable age determination (see \citealt{SNID}). It was also chosen to
include a sufficient number of intrinsically overluminous \sneia:
indeed, there are five such \sneia\ (SNe~1997br, 1998ab, 1999dq,
1999gp, and 2001eh) in this subsample, accounting for $\sim20$\% by
number of objects and spectra. For these specific tests, the templates
corresponding to the 22 \sneia\ in this subsample are temporarily
removed from the SNID database to avoid biasing the age determination. 

We deliberately restrict this subsample to ages between
$-10$ and +30~d from maximum light. Before $-10$ d, the number
of spectral templates in the SNID database drops rapidly, and the age
determination is inaccurate. Past +30 d, the spectra of \sneia\ do
not evolve as rapidly as around maximum light, and the age
determination is less precise \citep{SNID}.


\section{Aging Rate at High Redshifts}\label{sect:aging}

\subsection{Spectroscopic Data}

Our aging rate measurements at high redshifts are based on a sample of
35 spectra of 13 \sneia\ in the redshift range $0.28 \le z \le
0.62$. These include previously published data by the High-$Z$
Supernova Search Team (SN~1996bj, \citealt{Riess/etal:1997a}), the Supernova
Cosmology Project (SN~1997ex, \citealt{Foley/etal:2005}; SN~2001go,
\citealt{Lidman/etal:2005}), and the ESSENCE project (SNe~2002iz, b027, and
2003js, \citealt{Matheson/etal:2005}). For SN~2001go we present our own
reductions of the three epochs of spectroscopic data obtained from
the ESO Science Archive
Facility\footnote{http://www.eso.org/sci/archive/ .}, as only the first
spectrum was published by \cite{Lidman/etal:2005}. The spectra of
SN~04D2an (the highest-redshift \snia\ in this sample) were obtained
by members of the Supernova Legacy Survey (SNLS), and will be
published as part of a larger sample of SNLS data by St\'ephane Basa and
coworkers. SN~2006tk will be published alongside the complete ESSENCE
supernova dataset in the near future.

The other five \sneia\ (SNe~2006mk, 2006sc, 2007tg, 2007tt, 2007un) are
ESSENCE targets which were observed spectroscopically through two
dedicated ``Target-of-Opportunity'' programs at the ESO Very Large
Telescope\footnote{Programs 078.D-0383 and 080.D-0477; PI: Jesper
Sollerman.}. 

\begin{figure*}
\epsscale{1.1}
\plotone{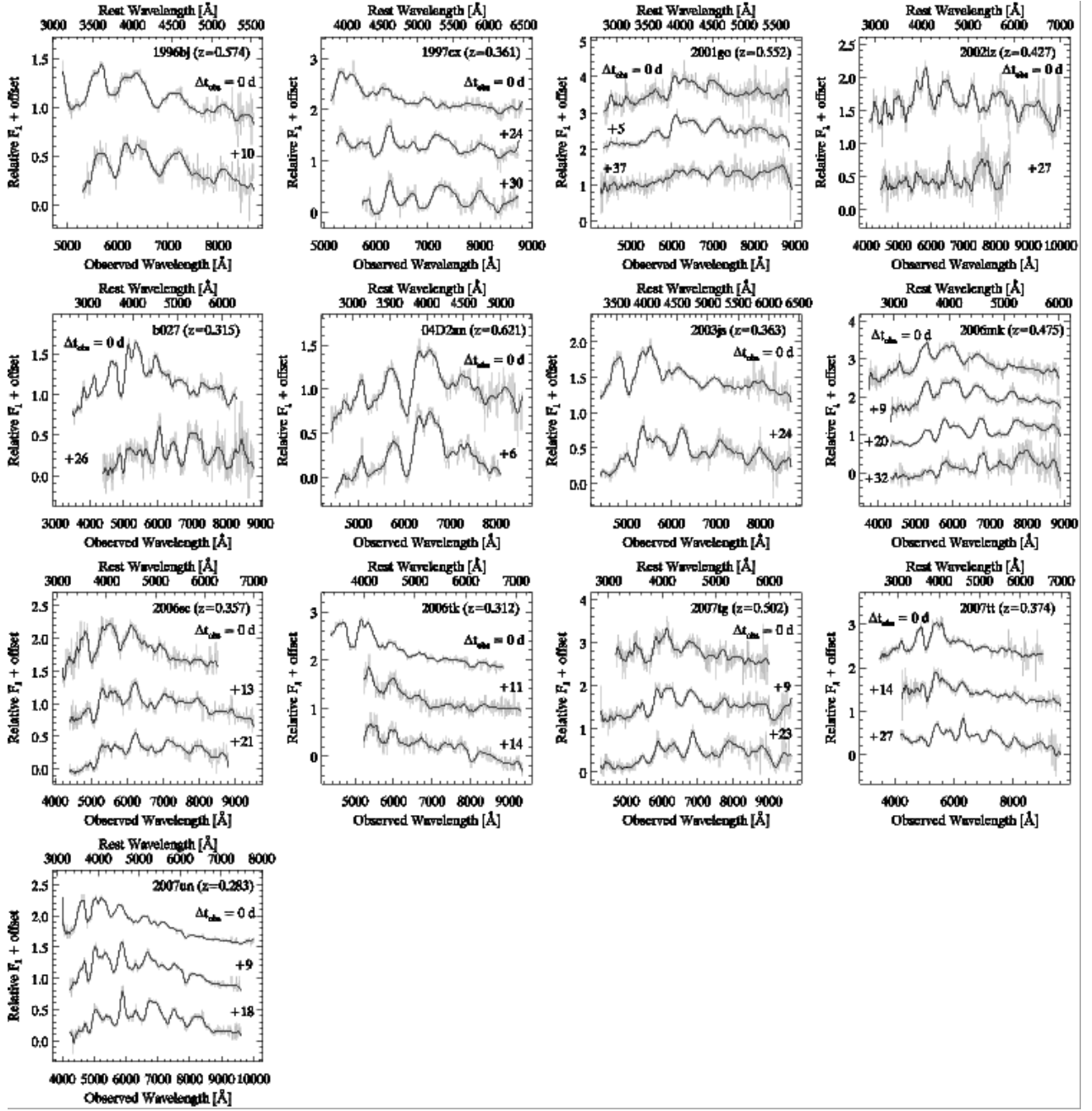}
\caption{\label{fig:spec}Multiepoch spectra of the 13 high-redshift \sneia\ used in
this study, binned to 10\,\AA\ (gray). The vertical offset
  between each spectrum is for clarity only, and does not reflect
  differences in flux density ($F_\lambda$;
  erg$~$s$^{-1}$cm$^{-2}$\AA$^{-1}$) between them. In each plot, the
age of the 
supernova increases downwards, and the observed time (in days) from
the first spectrum is indicated. Overplotted in black is a smoothed
version using the inverse-variance-weighted Gaussian algorithm of
\citet{Blondin/etal:2006}.}
\end{figure*}

Details on the instrumental setup and data reduction can
be found in the aforementioned references. The rest of the data will
be presented more thoroughly in a separate publication. All of the data
are shown in Fig.~\ref{fig:spec}. 

We use SNID to determine the redshift of each spectrum. The redshift
of a given supernova ($z_{\rm SN}$) is then 
computed as the error-weighted mean of the SNID redshifts ($z_{\rm
  SNID}$) for each of its spectra (Table~\ref{table:zsnid}). For ten
\sneia, we also have a redshift determination from the host
galaxy ($z_{\rm GAL}$). Comparison with the supernova redshift shows
an excellent agreement between the two measurements (better than
1\%). For the three remaining \sneia\ (SNe~b027, 2006tk, and 2007tg), only
the SNID redshift is available, but the
different redshift measurements for individual spectra all agree to
within 1$\sigma$, and we are confident about their accuracy. In what
follows, we will use the galaxy redshift when available for the age
and aging rate measurements. Given the excellent agreement between
$z_{\rm GAL}$ and $z_{\rm SN}$, this choice has negligible impact on
our results.

\begin{deluxetable}{lccc}
\tablewidth{0pt}
\tablecaption{\label{table:zsnid}Comparison of galaxy and supernova redshifts}
\tablehead{
\colhead{SN} &
\colhead{$z_{\rm GAL}$} & 
\colhead{$z_{\rm SN}$} & 
\colhead{$z_{\rm SNID}$} \\
\colhead{(1)} & 
\colhead{(2)} & 
\colhead{(3)} & 
\colhead{(4)}
}
\startdata
1996bj & 0.574   & 0.581 (0.005) & 0.580 (0.008) \\
       &         &               & 0.582 (0.008) \\
\hline
1997ex & 0.361   & 0.362 (0.002) & 0.362 (0.005) \\
       &         &               & 0.361 (0.004) \\
       &         &               & 0.362 (0.004) \\
\hline
2001go & 0.552   & 0.552 (0.005) & 0.552 (0.008) \\
       &         &               & 0.556 (0.008) \\
       &         &               & 0.550 (0.009) \\
\hline
2002iz & 0.427   & 0.425 (0.004) & 0.422 (0.006) \\
       &         &               & 0.428 (0.006) \\
\hline
b027   & \nodata & 0.315 (0.003) & 0.315 (0.006) \\
       &         &               & 0.315 (0.004) \\
\hline
2003js & 0.363   & 0.361 (0.003) & 0.359 (0.004) \\
       &         &               & 0.363 (0.006) \\
\hline
04D2an & 0.621   & 0.614 (0.006) & 0.608 (0.007) \\
       &         &               & 0.625 (0.011) \\
\hline
2006mk & 0.475   & 0.477 (0.003) & 0.479 (0.005) \\
       &         &               & 0.478 (0.007) \\
       &         &               & 0.474 (0.008) \\
       &         &               & 0.476 (0.006) \\
\hline
2006sc & 0.357   & 0.356 (0.004) & 0.355 (0.007) \\
       &         &               & 0.357 (0.007) \\
       &         &               & 0.356 (0.006) \\
\hline
2006tk & \nodata & 0.312 (0.003) & 0.312 (0.006) \\
       &         &               & 0.310 (0.003) \\
       &         &               & 0.315 (0.006) \\
\hline
2007tg & \nodata & 0.502 (0.004) & 0.503 (0.009) \\
       &         &               & 0.503 (0.008) \\
       &         &               & 0.502 (0.007) \\
\hline
2007tt & 0.374   & 0.376 (0.004) & 0.367 (0.008) \\
       &         &               & 0.379 (0.007) \\
       &         &               & 0.377 (0.005) \\
\hline
2007un & 0.283   & 0.285 (0.004) & 0.287 (0.006) \\
       &         &               & 0.285 (0.007) \\
       &         &               & 0.285 (0.005) 
\enddata
\tablenotetext{\ }{{\it Column headings:}
(1) SN name.
(2) Galaxy redshift (the typical error is $<0.001$).
(3) SN redshift, quoted as the error-weighted mean of the individual
  redshifts for each epoch.
(4) SNID redshift for each epoch, in order of increasing age.
}
\end{deluxetable}

\subsection{\label{sect:tdiff}Accuracy of Relative Age Determination}

An accurate determination of the rate of aging involves accurate
knowledge of age {\it differences}. In what follows we test how well
SNID determines differential ages using the subsample of 
low-redshift \sneia\ presented in Section~\ref{sect:snid}. While
the determination of {\it absolute} ages has no impact on the 
main result of this paper, we discuss their accuracy in
Appendix~\ref{sect:tcomp}.

We determine the rest-frame age ($t_{\rm spec}$) of each of the 145
spectra in the low-redshift subsample of 22 \sneia. We then compute
the absolute age difference ($\Delta t_{\rm spec}$)  between each
unique pair of spectra corresponding to a given supernova. This
amounts to 631 pairs. This age difference is then compared with the
absolute observer-frame age difference ($\Delta t_{\rm obs}$) for each
spectrum pair. Since $z \approx 0$ for this subsample, $\Delta
t_{\rm spec}$ can be directly compared to $\Delta t_{\rm obs}$ with
no correction for time dilation. Given the restriction to ages between
$-10$ and +30~d from maximum light in the low-redshift subsample,
$\Delta t_{\rm spec}$ (and hence $\Delta t_{\rm obs}$) is at most 40 d.

\begin{figure}
\epsscale{1.1}
\plotone{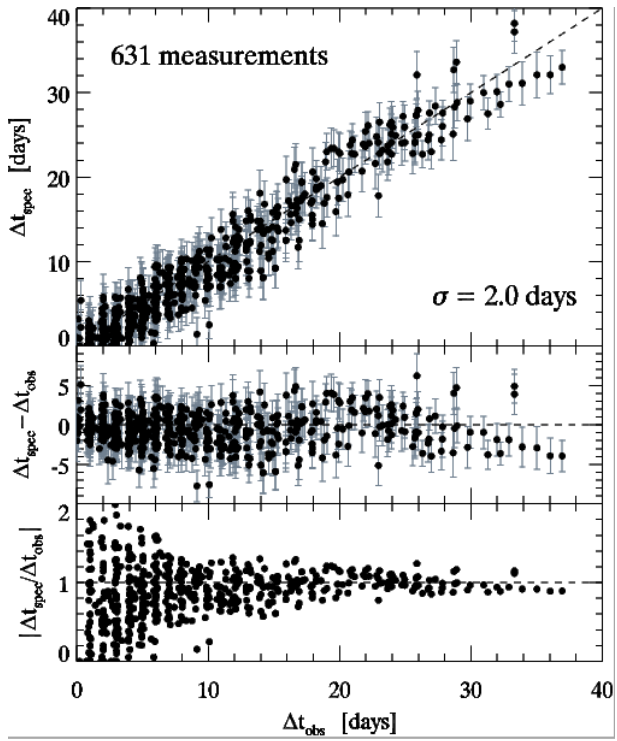}
\caption{\label{fig:tdiff}
Upper panel: Rest-frame age difference ($\Delta t_{\rm spec}$)
vs. observer-frame age difference ($\Delta t_{\rm obs}$) for each
spectrum pair for a given supernova. There are 631 such pairs, with a
dispersion of 2.0~d about the one-to-one correspondence.
Middle panel: Residuals in the upper panel vs. $\Delta t_{\rm obs}$.
Lower panel: Ratio of $\Delta t_{\rm spec}$ to $\Delta t_{\rm obs}$,
again vs. $\Delta t_{\rm obs}$. For $\Delta t_{\rm obs} > 6$ d, the
fractional difference is less than 20\%.
}
\end{figure}

The results are displayed in the upper panel of
Fig.~\ref{fig:tdiff}. There is good agreement between $\Delta t_{\rm
  spec}$ and $\Delta t_{\rm obs}$, with a dispersion of only 2.0 d
about the one-to-one correspondence. For $\Delta t_{\rm obs} \gtrsim
30$ d, however, SNID systematically underestimates the age
difference by $\sim 1.5$ d. This is more apparent in the plot of
residuals in the middle panel. it is mainly due to a systematic
overestimate of rest-frame ages $t_{\rm spec}\lesssim-7$~d from
maximum light, due to the lack of spectral templates in the SNID
database with similar ages (see Fig.~\ref{fig:thist} and
\citealt{SNID}).

The lower panel of Fig.~\ref{fig:tdiff} shows the absolute fractional
age difference vs. $\Delta t_{\rm obs}$. The quantity $|\Delta t_{\rm
  obs}/\Delta t_{\rm spec}|$ is a direct measure of the accuracy we can
achieve for the aging rate determination. As expected, the fractional
age difference decreases with increasing age difference. For $\Delta
t_{\rm obs} > 6$ d, this difference drops below 20\%. The
high-redshift data presented in the previous section span a sufficient
range of observer-frame age difference that the aging rate
determination is accurate. Note that the systematic underestimate of
the age difference for $\Delta t_{\rm obs} \gtrsim 30$~d results in
a negligible fractional difference.

\subsection{\label{sect:trate}Aging Rate Determination}

The rest-frame age of each high-redshift \snia\ spectrum ($t_{\rm
spec}$) is determined as outlined in Section~\ref{sect:snid}. In each
case, we fix the redshift to that determined in the previous section.
The results are displayed in Table~\ref{table:tdiff}, along with the
corresponding observed date of each spectrum ($t_{\rm obs}$). However,
since the aging rate determination depends on age {\it differences} (see
previous Section), we also report the observer-frame and rest-frame
age from the first spectrum, respectively denoted $\Delta t_{\rm obs}$
and $\Delta t_{\rm spec}$ in Table~\ref{table:tdiff}.

\begin{figure*}
\epsscale{1.1}
\plotone{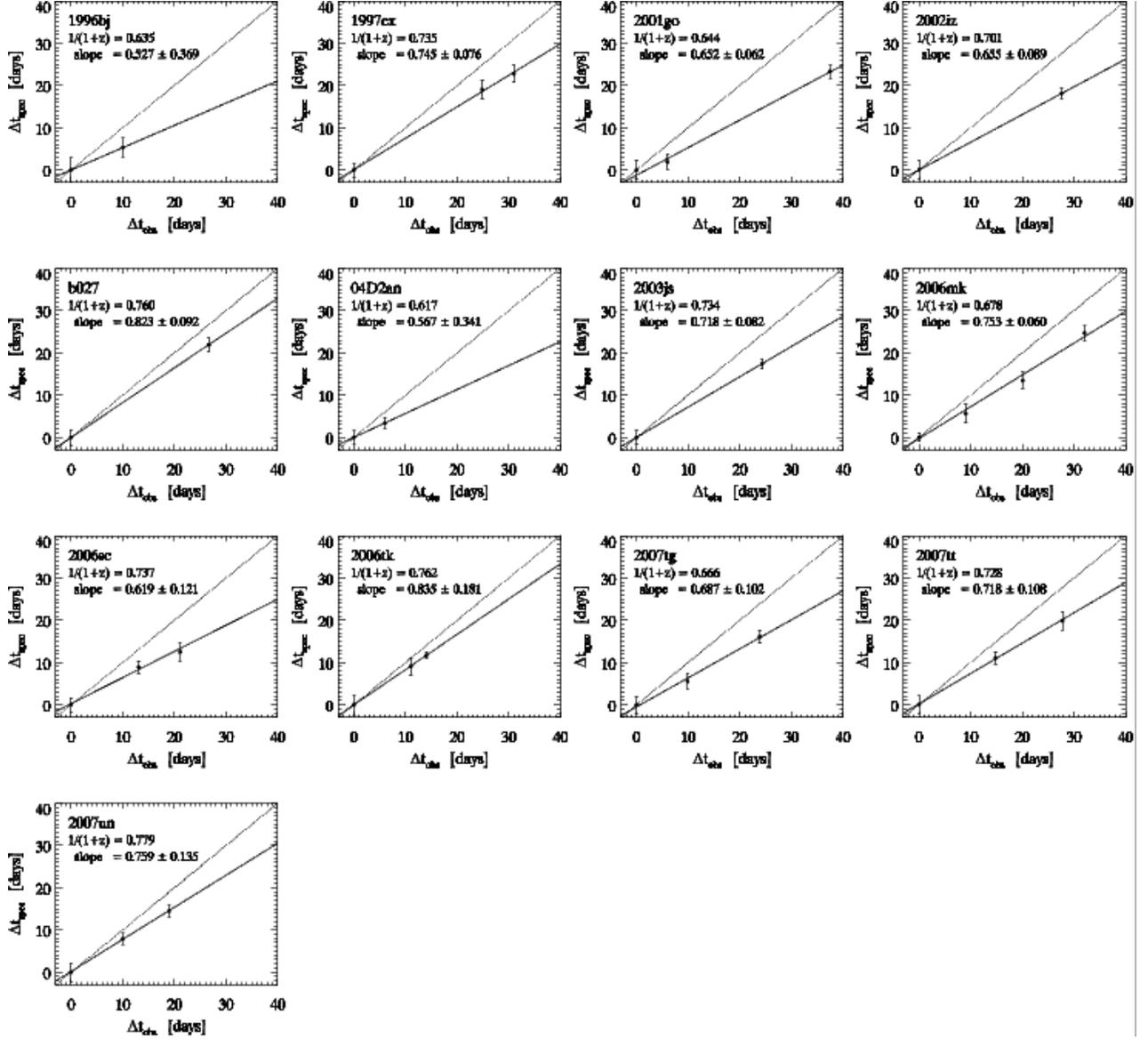}
\caption{\label{fig:tdilfig}
 Comparison of rest-frame ($\Delta t_{\rm spec}$) and observer-frame
 ($\Delta t_{\rm obs}$) time from the first spectrum, for each of the
 13 high-redshift \sneia\ in our sample. The abscissa and ordinate
 ranges are both set to [$-$3,+40]~d in all cases. The slope of
the best-fit line (solid line) gives a measurement of the
apparent aging rate of the supernova, which is compared to the
expected $1/(1+z)$ value. The dotted line in each plot
  corresponds to $\Delta t_{\rm spec}=\Delta t_{\rm obs}$.
}
\end{figure*}

We can then trivially compute the aging rate for each supernova. This
is simply done through a least-squares fit of a line to $\Delta t_{\rm
 spec}$ versus $\Delta t_{\rm obs}$. The slope of the line is a
measure of the aging rate, which should equal $1/(1+z)$ in an
expanding universe (see Appendix~\ref{sect:math}). Were there no time
dilation, the aging rate would equal one. The results are displayed in
Fig.~\ref{fig:tdilfig}.

\begin{deluxetable}{lrrrr}
\tablewidth{0pt}
\tablecaption{\label{table:tdiff}Observer-frame and rest-frame age differences}
\tablehead{
\colhead{SN} &
\colhead{$t_{\rm obs}$} & 
\colhead{$t_{\rm spec}$} & 
\colhead{$\Delta t_{\rm obs}$} & 
\colhead{$\Delta t_{\rm spec}$} \\
\colhead{(1)} & 
\colhead{(2)} & 
\colhead{(3)} & 
\colhead{(4)} & 
\colhead{(5)}
}
\startdata
1996bj &  367.99 &$-$2.2 (3.0) &  0.00 &   0.0 (3.0) \\
       &  378.04 &   3.1 (2.2) & 10.05 &   5.3 (2.2) \\
\hline
1997ex &  815.08 &$-$1.6 (1.6) &  0.00 &   0.0 (1.6) \\
       &  839.96 &  17.4 (2.1) & 24.88 &  19.0 (2.1) \\
       &  846.03 &  21.2 (2.0) & 30.95 &  22.8 (2.0) \\
\hline
2001go & 2021.70 &   7.9 (2.3) &  0.00 &   0.0 (2.3) \\
       & 2027.69 &   9.8 (1.7) &  5.99 &   1.9 (1.7) \\
       & 2059.17 &  31.2 (1.6) & 37.47 &  23.3 (1.6) \\
\hline
2002iz & 2586.95 &$-$0.5 (2.2) &  0.00 &   0.0 (2.2) \\
       & 2614.57 &  17.6 (1.2) & 27.62 &  18.1 (1.2) \\
\hline
b027   & 2589.95 &$-$3.5 (1.8) &  0.00 &   0.0 (1.8) \\
       & 2616.57 &  18.4 (1.6) & 26.62 &  21.9 (1.6) \\
\hline
2003js & 2942.46 &$-$4.9 (1.6) &  0.00 &   0.0 (1.6) \\
       & 2966.71 &  12.5 (1.2) & 24.25 &  17.4 (1.2) \\
\hline
04D2an & 3026.20 &$-$2.5 (1.6) &  0.00 &   0.0 (1.6) \\
       & 3032.20 &   0.9 (1.3) &  6.00 &   3.4 (1.3) \\
\hline
2006mk & 4031.71 &$-$6.2 (1.0) &  0.00 &   0.0 (1.0) \\
       & 4040.72 &$-$0.6 (2.2) &  9.01 &   5.6 (2.2) \\
       & 4051.77 &   7.3 (1.9) & 20.06 &  13.5 (1.9) \\
       & 4063.78 &  18.5 (1.8) & 32.07 &  24.7 (1.8) \\
\hline
2006sc & 4063.58 &   0.9 (1.6) &  0.00 &   0.0 (1.6) \\
       & 4076.65 &   9.8 (1.4) & 13.07 &   8.9 (1.4) \\
       & 4084.68 &  13.4 (2.2) & 21.10 &  12.5 (2.2) \\
\hline
2006tk & 4089.57 &$-$8.8 (2.4) &  0.00 &   0.0 (2.4) \\
       & 4100.57 &   0.3 (2.0) & 11.00 &   9.1 (2.0) \\
       & 4103.59 &   2.9 (0.9) & 14.02 &  11.7 (0.9) \\
\hline
2007tg & 4381.75 &$-$6.1 (2.0) &  0.00 &   0.0 (2.0) \\
       & 4391.65 &$-$0.5 (1.8) &  9.90 &   5.6 (1.8) \\
       & 4405.61 &  10.0 (1.5) & 23.86 &  16.1 (1.5) \\
\hline
2007tt & 4415.81 &$-$5.0 (2.1) &  0.00 &   0.0 (2.1) \\
       & 4430.65 &   6.1 (1.4) & 14.84 &  11.1 (1.4) \\
       & 4443.58 &  14.9 (2.2) & 27.77 &  19.9 (2.2) \\
\hline
2007un & 4441.61 &   3.2 (2.2) &  0.00 &   0.0 (2.2) \\
       & 4451.60 &  11.1 (1.3) &  9.99 &   7.9 (1.3) \\
       & 4460.58 &  17.7 (1.4) & 18.97 &  14.5 (1.4) 
\enddata
\tablenotetext{\ }{{\it Column headings:}
(1) SN name.
(2) Julian date ($JD$) minus 2,450,000 at midpoint of observation.
(3) SN rest-frame age in days from maximum light, derived from the
  cross-correlation with spectral templates using SNID.
(4) Observer-frame days from first spectrum.
(5) Rest-frame days from first spectrum.
}
\end{deluxetable}

\begin{figure}
\epsscale{1.1}
\plotone{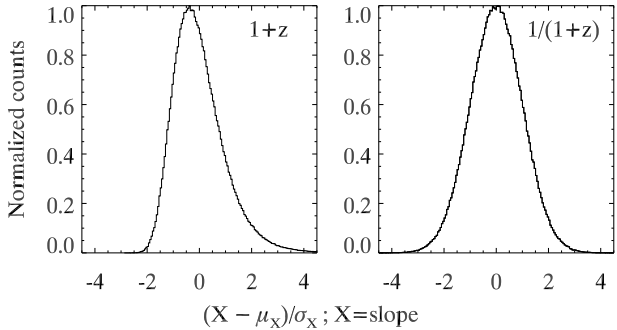}
\caption{\label{fig:slopemc}
  Monte Carlo results illustrating the advantage of working in
  $1/(1+z)$ space for time-dilation measurements. Solid lines:
  recovered slope (in standard deviations from the mean, $\mu$) in
  $(1+z)$ (left) and $1/(1+z)$ space (right),  using
  the SN age errors reported by \cite{Foley/etal:2005}. The
  distribution is highly non-Gaussian in the former case. 
}
\end{figure}

We note that comparing the inverse of the slope in
Fig.~\ref{fig:tdilfig} (denoted ``age factor'' by
\citealt{Foley/etal:2005}) and $(1+z)$ leads to asymmetric errors.
The errors on the ``age
factor'' [$\equiv (1+z)$] become highly non-Gaussian when the uncertainties of
the individual age measurements are large ($\gtrsim 1$ d, as is the
case in this paper, and in \citealt{Foley/etal:2005} for SN 1997ex), whereas the errors
on the aging rate [$\equiv 1/(1+z)$] are always Gaussian. This is
illustrated in Fig.~\ref{fig:slopemc} using a Monte Carlo simulation
of the age measurements for SN~1997ex presented by
\cite{Foley/etal:2005}. Using the same errors on the individual age
measurements, the distribution of the slope measurements is highly
non-Gaussian in $(1+z)$ space, while it is normally distributed in
$1/(1+z)$ space. 

The individual aging rate measurements presented here alone reject
models that predict no time dilation at a high significance (up to 
$\sim6\sigma$), and all (except for SN~2006mk) are within $1\sigma$ of
the expected $1/(1+z)$ factor. In the next section we combine all
aging rate measurements (including those for the low-redshift sample)
to test each hypothesis more thoroughly.


\section{Testing the $1/(1+\lowercase{z})$ Time-Dilation Hypothesis}\label{sect:tdil}

We have determined the aging rate for the subsample of 22 low-redshift
\sneia. We combine these aging rates with those determined for the 13
high-redshift \sneia\ of our sample (see Table~\ref{table:tdil}) to
test the $1/(1+z)$ time-dilation hypothesis. As noted in
Section~\ref{sect:tdiff} and Appendix~\ref{sect:tcomp},
these measurements rely on a database of
\snia\ spectra whose ages have already been corrected for the expected
$1/(1+z)$ time-dilation factor. However, since all \sneia\ in the SNID
database are at redshifts $z\le0.05$, the correction is small
($\lesssim 1$ d) and has a negligible impact on the aging rate
measurements.

\begin{deluxetable}{lccc}
\tablewidth{0pt}
\tablecaption{\label{table:tdil}Aging rate measurements}
\tablehead{
\colhead{SN} &
\colhead{$z$} & 
\colhead{$1/(1+z)$} & 
\colhead{Aging rate}
}
\startdata
\multicolumn{4}{c}{Low redshift ($z < 0.04$)} \\
\hline
1981B  & 0.006 & 0.994 & 1.099 (0.071) \\
1989B  & 0.002 & 0.998 & 1.036 (0.079) \\
1992A  & 0.006 & 0.994 & 1.040 (0.077) \\
1994D  & 0.002 & 0.998 & 1.011 (0.152) \\
1996X  & 0.007 & 0.993 & 0.886 (0.116) \\
1997br & 0.007 & 0.993 & 1.120 (0.149) \\
1998V  & 0.018 & 0.983 & 1.047 (0.166) \\
1998ab & 0.027 & 0.974 & 0.987 (0.151) \\
1998dm & 0.007 & 0.993 & 0.778 (0.119) \\
1998eg & 0.025 & 0.976 & 0.883 (0.085) \\
1999cl & 0.008 & 0.992 & 0.870 (0.122) \\
1999dq & 0.014 & 0.986 & 0.928 (0.055) \\
1999ej & 0.014 & 0.986 & 1.147 (0.159) \\
1999gp & 0.027 & 0.974 & 0.839 (0.060) \\
2000fa & 0.021 & 0.979 & 0.884 (0.076) \\
2001V  & 0.015 & 0.985 & 0.973 (0.072) \\
2001eh & 0.037 & 0.964 & 1.148 (0.107) \\
2001ep & 0.013 & 0.987 & 0.925 (0.086) \\
2002ha & 0.014 & 0.986 & 0.973 (0.134) \\
2003cg & 0.004 & 0.996 & 1.006 (0.131) \\
2003du & 0.006 & 0.994 & 0.909 (0.091) \\
2006lf & 0.013 & 0.987 & 0.941 (0.110) \\
\hline
\multicolumn{4}{c}{High redshift ($z > 0.2$)} \\
\hline
1996bj & 0.574 & 0.635 & 0.527 (0.369) \\
1997ex & 0.361 & 0.735 & 0.745 (0.076) \\
2001go & 0.552 & 0.644 & 0.652 (0.062) \\
2002iz & 0.427 & 0.701 & 0.655 (0.089) \\
b027   & 0.315 & 0.760 & 0.823 (0.092) \\
2003js & 0.363 & 0.734 & 0.718 (0.082) \\
04D2an & 0.621 & 0.617 & 0.567 (0.341) \\
2006mk & 0.475 & 0.678 & 0.753 (0.060) \\
2006sc & 0.357 & 0.737 & 0.619 (0.121) \\
2006tk & 0.312 & 0.762 & 0.835 (0.181) \\
2007tg & 0.502 & 0.666 & 0.687 (0.102) \\
2007tt & 0.374 & 0.728 & 0.718 (0.108) \\
2007un & 0.283 & 0.779 & 0.759 (0.135) 
\enddata
\end{deluxetable}

\begin{figure*}
\epsscale{1.1}
\plotone{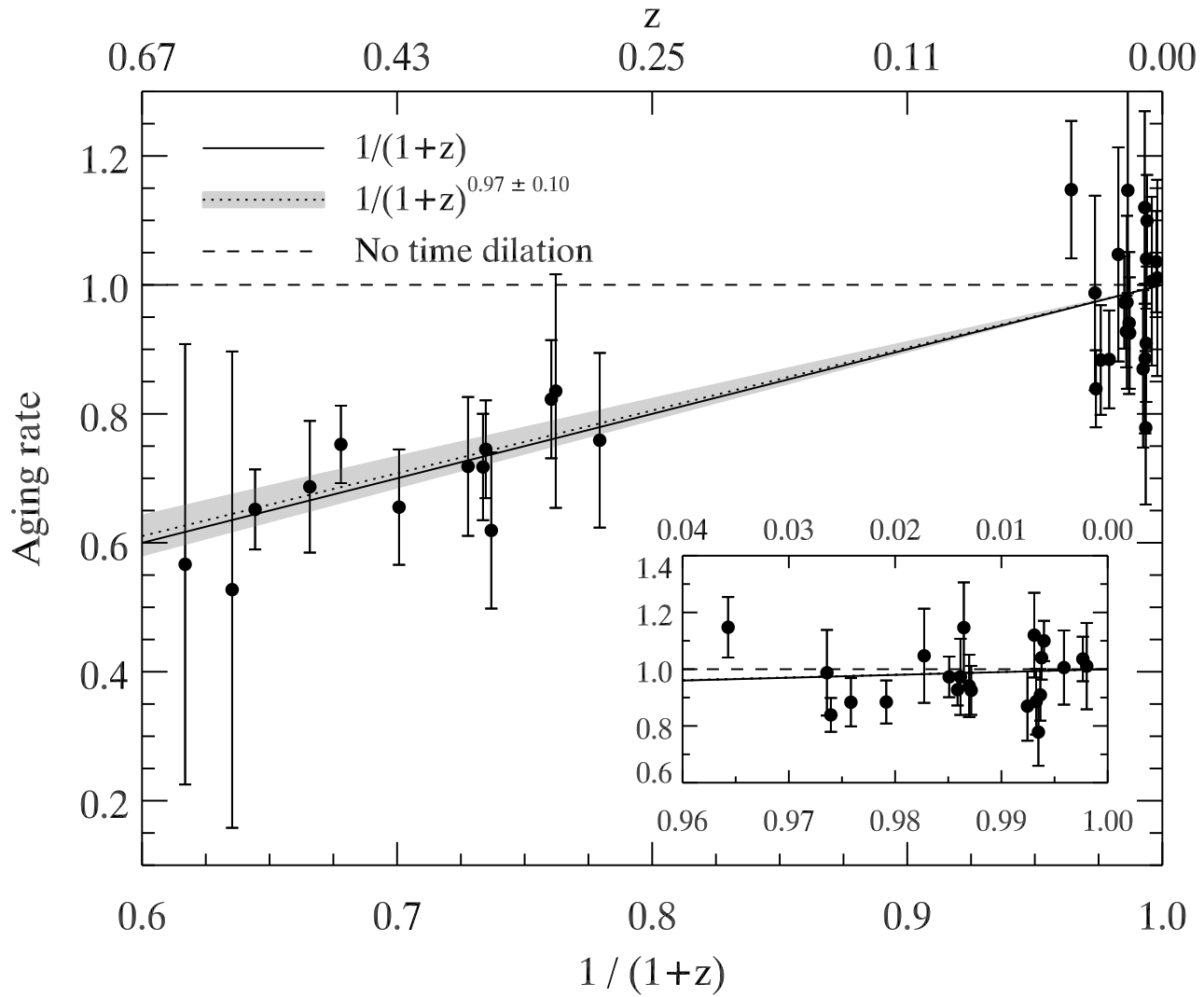}
\caption{\label{fig:tdil}
Apparent aging rate versus $1/(1+z)$ for the 13 high-redshift ($0.28 \le
z \le 0.62$) and 22 low-redshift ($z < 0.04$) \sneia\ in our
sample. Overplotted are the expected $1/(1+z)$ time dilation (solid
line) and the best-fit $1/(1+z)^b$ model (with $b=0.97\pm0.10$; dotted
line and gray area). The dashed line corresponds to no time dilation,
as expected in the tired-light model --- clearly inconsistent with the
data. The inset shows a close-up view of the low-redshift
sample. These data are summarized in Table~\ref{table:tdil}.
}
\end{figure*}

All aging rate measurements are shown in Fig.~\ref{fig:tdil}. The
solid line shows the expected $1/(1+z)$ time-dilation factor, while
the dashed line represents the ``tired light'' hypothesis of
\citet{Zwicky:1929}. According to this hypothesis, photons lose energy
as they interact with matter and other photons in a static
universe. The energy loss is proportional to the distance from the
source, and causes a redshift in spectra as in an expanding
universe. However, this hypothesis does not predict a
time-dilation effect, and so the aging rate should equal one for all
redshifts.

As expected, the measurement of a time-dilation effect is more obvious
at larger redshift, and the precision improves as the number and time
span of spectra for each supernova increases. This latter effect
explains why the aging rate measurements for SN~1996bj ($z=0.574$) and
SN~04D2an ($z=0.621$) have a large associated error --- despite being
the two highest-redshift \sneia\ in our sample, since
only two spectra separated by $\sim10$ (for SN~1996bj) and $\sim6$
(for SN~04D2an) observer-frame days are available.

A simple $\chi^2$ analysis is sufficient to confirm what the eye sees:
the hypothesis of no time dilation is not a good fit to the data
($\chi^2=150.3$ for 35 degrees of freedom; see
Table~\ref{table:modelcomp}), with a goodness-of-fit of $\sim0\%$
(defined as GoF$=\Gamma(\nu/2,\chi^2/2)/\Gamma(\nu/2)$, where
$\Gamma(\nu/2,\chi^2/2)$ is the incomplete gamma function and $\nu$ is
the number of degrees of freedom) --- namely, a null probability of
obtaining data that are a worse fit to the model, assuming that the
model is indeed correct. The expected $1/(1+z)$ time-dilation factor,
on the other hand, yields a good fit to the data ($\chi^2=27.0$
for 35 degrees of freedom), with GoF$=83.2\%$, and is largely
favored over the null hypothesis of no time dilation ($\Delta \chi^2
\approx 123$).

\begin{deluxetable*}{lrrrrrrrrr}
\tablewidth{0pt}
\tablecaption{\label{table:modelcomp}Time-dilation model comparison}
\tablehead{
 &
\multicolumn{3}{c}{All SNe} &
\multicolumn{3}{c}{High-redshift SNe only} & 
\multicolumn{3}{c}{Low-redshift SNe only} \\
\cline{2-4} \cline{5-7} \cline{8-10}
Model\tablenotemark{a} & 
$\chi^2$/dof & GoF & $\Delta$AIC & 
$\chi^2$/dof & GoF & $\Delta$AIC & 
$\chi^2$/dof & GoF & $\Delta$AIC \\
 & & (\%) & & & (\%) & & & (\%) & 
}
\startdata
$1/(1+z)$       &  27.0/35 &  83.2 &   0 &   3.6/13 &  99.5 &   0 &  23.4/22 &  38.2 &  0 \\
$1/(1+z)^b$     &  26.9/34 &  80.2 &   1 &   3.4/12 &  99.2 &   1 &  20.3/21 &  50.0 &$-$1\\
tired light     & 150.3/35 &   0.0 & 123 & 123.4/13 &   0.0 & 119 &  26.9/22 &  21.4 &  3 
\enddata
\tablenotetext{a}{The best-fit values for the $b$ exponent in the
      second model are as follows. 
      All SNe: $b=0.97\pm0.10$;
      high-redshift SNe only: $b=0.95\pm0.10$; 
      low-redshift SNe only: $b=3.18\pm1.28$}
\end{deluxetable*}

This result holds (and in fact improves)
when we consider only the high-redshift sample (see
Table~\ref{table:modelcomp}). 
This works because the $z=0$ end of the aging rate versus redshift
relation (Fig.~\ref{fig:tdil}) is fixed to unity by theory, so the
low-redshift sample is not needed to anchor the theoretical curve at
$z\approx 0$ (although it is still used to calibrate the $t_{\rm spec}$
measurement).
The low-redshift data alone do not enable us to distinguish between
the two hypotheses, since the impact of time dilation is small at such
low redshifts.

\begin{figure}
\epsscale{1.1}
\plotone{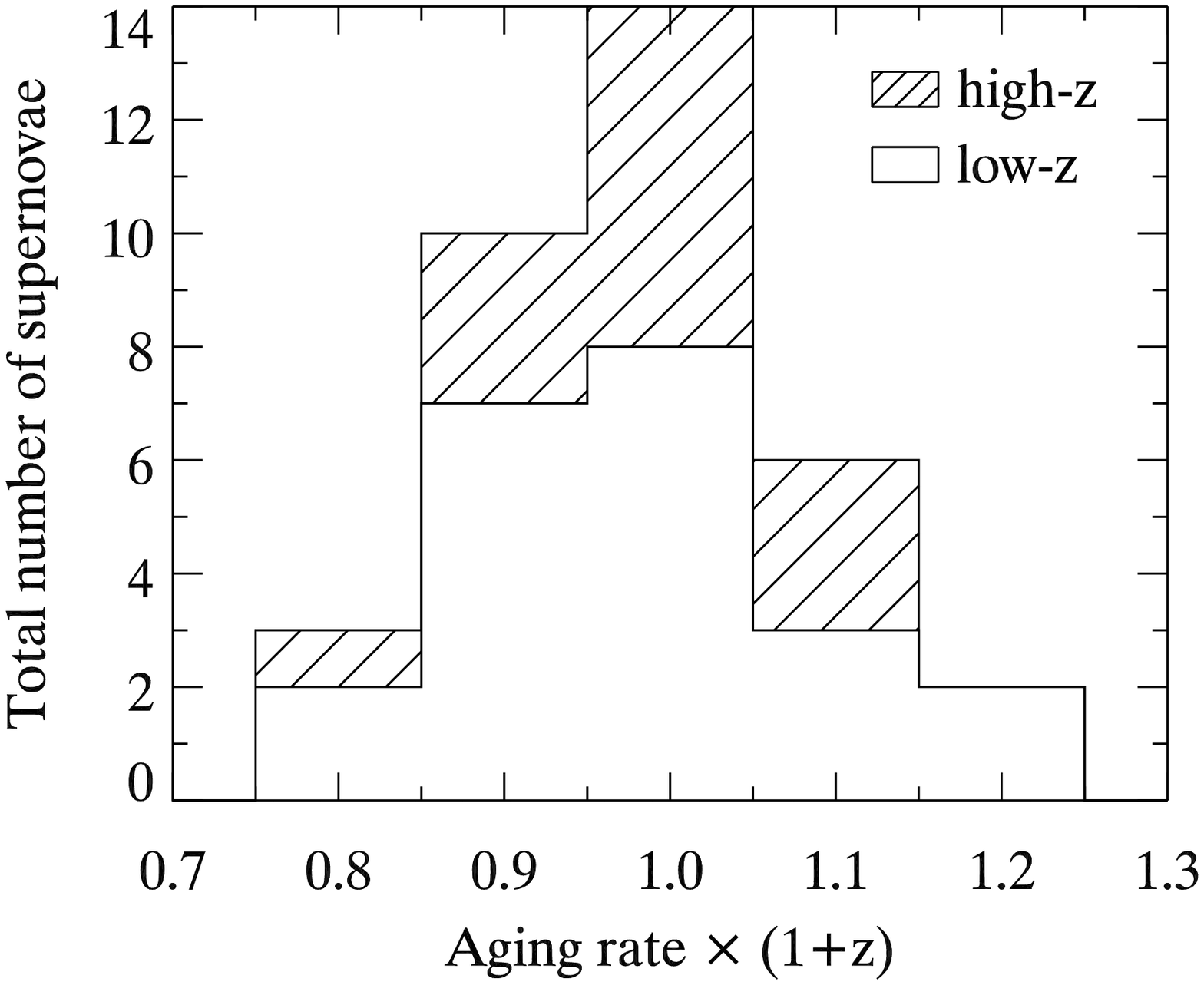}
\caption{\label{fig:tdil_ratio}
Ratio of the aging rate to $1/(1+z)$ for all \sneia\ in
Fig.~\ref{fig:tdil}. Both the low-redshift (open histogram) and
high-redshift (hatched histogram) samples are shown.
}
\end{figure}

In Fig.~\ref{fig:tdil_ratio} a different view of Fig.~\ref{fig:tdil}
shows the distributions of the ratio between the aging rate and
$1/(1+z)$ for both the low-redshift (open histogram) and high-redshift
(hatched histogram) samples. Both distributions are within $\sim20\%$
of a unit ratio, again validating the hypothesis of time dilation over
a large redshift range. The apparent bias to lower values of the
ratio for the low-redshift sample is not statistically significant, as
the mean error on the aging rate is of order one bin size
($\lesssim 0.1$; see Table~\ref{table:tdil}).

In what follows we test whether the data favor a nonlinear dependence
of the aging rate on redshift, namely

\begin{equation}
\label{eqn:power}
{\rm aging\ rate} = \frac{1}{(1+z)^b},
\end{equation}

\noindent
where $b$ is a free parameter. While Eq.~[\ref{eqn:power}] 
satisfies the same zero point as the two previous hypotheses (aging
rate equal to 1 at $z=0$), no model actually predicts such a
dependence of the aging rate on redshift. Nonetheless, small
deviations from the expected $1/(1+z)$ factor would have profound
implications for our assumption of FLRW cosmology.

Again, we performed a least-squares fit to the entire sample, and also
to the individual high- and low-redshift samples (see
Table~\ref{table:modelcomp}). 
The data constrain the $b$ exponent to 10\%
($1\sigma$), and yield $b=0.97\pm0.10$ for the entire sample
(Fig.~\ref{fig:tdil}; dotted line and gray region) and
$b=0.95\pm0.10$ for the high-redshift sample. As expected, the
low-redshift sample alone is insufficient to constrain the free
parameter ($b=3.18\pm1.28$). Nonetheless, the samples that include the 
high-redshift objects have a best-fit value for $b$ that is consistent
with $b=1$, and thus with the expected $1/(1+z)$ time-dilation
factor.

Since this model has an additional free parameter, it is
instructive to use information criteria to compare it to the simple
$1/(1+z)$ prediction.
These model comparison statistics favor models that yield a good fit
to the data with fewer parameters. 
As in \cite{Davis/etal:2007}, we use the 
Akaike information criterion (AIC; \citealt{AIC}). For Gaussian errors
(which is the case here, see Section~\ref{sect:trate}), 
this criterion can be expressed as

\begin{equation}
\label{eqn:aic}
{\rm AIC} = \chi^2 + 2k,
\end{equation}

\noindent
where $k$ is the number of free parameters
\citep{Davis/etal:2007}. Comparison of
models simply involves computing the difference in
AIC ($\Delta$AIC) with respect to the model with the lowest value for
this criterion. A difference in
AIC of 2 is considered positive evidence against the model with the
higher AIC, whereas a difference of 6 is considered strong evidence
\citep{Liddle:2004,Davis/etal:2007}. In the models considered here (see
Table~\ref{table:modelcomp}), the expected $1/(1+z)$ time-dilation
model has the lowest
AIC (although this is not true for the low-redshift sample), and we
compute AIC differences with respect to that model. 

With $\Delta {\rm AIC}=1$, we conclude that the information
criteria do not provide positive evidence against a $1/(1+z)^b$
dependence of the aging rate. The $\chi^2$ per degree of freedom is
also satisfactory for the samples that include the high-redshift
objects.

The two other models considered previously have no free parameters
($k=0$ in Eq.~[\ref{eqn:aic}]), hence $\Delta {\rm
 AIC}=\Delta\chi^2$, and the information criterion is reduced to a
  simple $\chi^2$ test.


\section{Conclusion}\label{sect:ccl}

We have presented 35 spectra of 13 high-redshift ($0.28 \le z \le
0.62$) \sneia, which include previously unpublished data from the
ESSENCE and SNLS projects and from our own dedicated program at the
ESO Very Large Telescope. Given the rapid and predictable evolution of
\snia\ spectral features with age, as well as the relative homogeneity
of \snia\ spectra at a given age, one is able to determine the 
(rest-frame) age of a single spectrum with a typical accuracy of 1--3
d
\citep{Riess/etal:1997a,Foley/etal:2005,Hook/etal:2005,Howell/etal:2005,SNID}.

Using the Supernova Identification (SNID) code of \citet{SNID}, we
determine the ages of each spectrum in the supernova
rest frame. Comparison with the observed time difference between the
spectra yields an apparent aging rate consistent
with $1/(1+z)$, as expected in a homogeneous and isotropic expanding
universe.
Moreover, the data unambiguously rule out the ``tired 
light'' hypothesis \citep{Zwicky:1929} in which photons lose energy as
they interact with matter and other photons in a static universe.

The fact that the age determination is so accurate over
a large redshift range also shows that the deviations between
spectra of low- and high-redshift \sneia\ in our sample are small.

We also test for alternate dependencies of the
aging rate on redshift, namely $1/(1+z)^b$, although these are not
predicted by any model. Whether we consider the entire sample or only
the high-redshift sample, the best-fit value for the $b$
exponent is consistent with $b=1$, and thus with the expected
$1/(1+z)$ factor.

That these data provide a confirmation of the time-dilation factor
expected in an expanding universe should be of no
surprise. Nonetheless, previous use of \snia\ light curves to test
this hypothesis \citep{SN1995K,Goldhaber/etal:2001} are
prone to the spread in intrinsic light-curve widths and its
possible variation with redshift (which includes selection effects;
see Section~\ref{sect:intro}).

The data presented here are unique in that they enable the most direct
test of the $1/(1+z)$ time-dilation hypothesis over a larger redshift
range than has yet been performed. This hypothesis is favored
beyond doubt over models that predict no time dilation. With more
data, the focus will shift to testing more thoroughly the alternative
$1/(1+z)^b$ dependence of the aging rate on redshift. Any significant
deviation from $b=1$ would have a profound impact on our assumption of
a FLRW cosmology to describe the universal expansion.


\begin{acknowledgments}
The authors wish to thank the Supernova Legacy Survey (SNLS)
collaboration, and in particular St\'ephane Basa and Tianmeng Zhang,
for providing spectra of supernova 04D2an prior to publication.
The ESSENCE survey is supported by the US National Science Foundation
(NSF) through grants AST 04-43378 and AST 05-07475. Support for
supernova research at Harvard University, including the CfA Supernova
Archive, is provided in part by NSF
grant AST 06-06772. The Dark Cosmology Centre is funded by the Danish
National Research Foundation. 
A.~C. acknowledges the support of CONICYT, Chile, under grants FONDECYT 
1051061 and FONDAP 15010003.
T.~M.~D. appreciates the support of the Villum Kann Rasmussen Fonden. 
A.~V.~F. is grateful for the support of NSF grant AST 06-07894.
G.~P. acknowledges the support of the Proyecto FONDECYT 3070034.
\end{acknowledgments}

{\it Facilities:}
\facility{VLT:Kueyen (FORS1)},
\facility{VLT:Antu (FORS2)},
\facility{Gemini:South (GMOS)},
\facility{Gemini:Gillete (GMOS)},
\facility{Keck:I (LRIS)},
\facility{Keck:II (DEIMOS, ESI)},
\facility{Magellan:Baade (IMACS)},
\facility{Magellan:Clay (LDSS3)}.


\begin{appendix}
\section{Time Dilation in an Expanding Universe}\label{sect:math}

In a homogeneous, isotropic expanding universe, the interval $d\tau$
between two space-time events is given by the Robertson-Walker (RW)
metric
\citep{Robertson:1935,Robertson:1936a,Robertson:1936b,Walker:1936},

\begin{equation}
\label{eqn:flrw}
d\tau^2 = c^2dt^2 - a^2(t) 
\left[ \frac{dr^2}{1-kr^2} + r^2(d\theta^2+\sin^2\theta d\phi^2) \right],
\end{equation}

\noindent
where $c$ is the speed of light in vacuum, $t$ is the cosmic time,
$(r,\theta,\phi)$ are the comoving spatial coordinates, $k$ is the
curvature parameter, and $a(t)$ is the dimensionless scale factor. In
what follows we assume the present-day value of the dimensionless
scale factor $a_0 = 1$.

Photons travel along null geodesics ($d\tau^2=0$). In what follows we
consider radial null rays only ($d\theta=d\phi=0$). For a photon
emitted at time $t_1$ from an object located at
$(r_1,\theta_1,\phi_1)$ and observed at time $t_0$,
Eq.~[\ref{eqn:flrw}] implies

\begin{equation}
\label{eqn:t}
\int_{t_1}^{t_0} \frac{cdt}{a(t)} = \int_0^{r_1} \frac{dr}{\sqrt{1-kr^2}}
\equiv f(r_1).
\end{equation}

\noindent
Here we assume that the object from which the photon was emitted
has constant coordinates $(r_1,\theta_1,\phi_1)$ such that $f(r_1)$,
also known as the comoving distance, is time independent. Thus, for a
photon emitted at time $t_1+\delta t_1$ and observed at time
$t_0+\delta t_0$, Eq.~[\ref{eqn:flrw}] also implies

\begin{equation}
\label{eqn:dt}
\int_{t_1+\delta t_1}^{t_0 + \delta t_0} \frac{cdt}{a(t)} = f(r_1).
\end{equation}

\noindent
For small $\delta t_1$ (and hence small $\delta t_0$), the rate of
change of the scale factor remains roughly constant and
Eqs.~[\ref{eqn:t}] and [\ref{eqn:dt}] imply

\begin{equation}
\frac{\delta t_0}{a_0} = \frac{\delta t_1}{a(t_1)}.
\end{equation}

\noindent
Hence, a light signal emitted with frequency $\nu_1$ will reach us with
frequency $\nu_0$ such that

\begin{equation}
\frac{\nu_0}{\nu_1} = \frac{\delta t_1}{\delta t_0} = \frac{a(t_1)}{a_0}.
\end{equation}

Using the standard definition of redshift, $z = (\l_0-\l_1)/\l_1 =
\nu_1/\nu_0 - 1$, we obtain a relationship between observed and
rest-frame time intervals in a RW metric as a function of redshift $z$:

\begin{equation}
\label{eqn:dtfrac}
\frac{\delta t_0}{\delta t_1} = 1+z.
\end{equation}

\noindent
A supernova at redshift $z$ will thus appear to age $(1+z)$ times more
slowly with respect to a local event at $z \approx 0$.

The prediction of time dilation proportional to $(1+z)$ is generic to
expanding universe models, whether the underlying theory be general
relativity (e.g., the Friedmann-Lema\^itre-Robertson-Walker universe),
special relativity (e.g., the Milne Universe), or Newtonian expansion.
A point of confusion can occur in the special relativistic case for
which the well-known time-dilation factor is given by

\bea 
\gamma_{\rm SR} & = & \left[1-\left(\frac{v}{c}\right)^2\right]^{-1/2} \\
 & = & \frac{1}{2} \left(1+z+\frac{1}{1+z}\right), 
\label{eq:SRtimedilation}
\eea

\noindent
which evidently differs from $(1+z)$.  Thus it might be assumed that a
special relativistic expansion can be distinguished from the FLRW
universe using a time-dilation test\footnote{In fact, such an
erroneous assumption was made by one of the current authors of
\cite{Davis/Lineweaver:2004}.}.
 
This is not the case.  
Special relativistic expansion of the  
universe assumes there is an inertial frame that extends to infinity
(impossible in the non-empty
general relativistic picture) and that the expansion involves objects
moving through this inertial frame.  The time-dilation factor from
Eq.~[\ref{eq:SRtimedilation}] relates the proper time in the moving
emitter's inertial frame ($\delta t_1$) to the proper time in the
observer's inertial frame ($\delta t_0$).   To measure this time
dilation the observer has to set up a set of synchronized clocks (each
at rest in the observer's inertial frame) and take readings of the
emitter's proper time as the emitter moves past each synchronized
clock.  The readings show that the emitter's clock is time dilated
such that $\delta t_0 = \gamma_{\rm SR}\delta t_1$.

We do not have this set of synchronized clocks at our disposal when
we measure time dilation of supernovae in an expanding universe
and therefore Eq.~[\ref{eq:SRtimedilation}] is not the time dilation we
observe.  We must also take into account an extra time-dilation factor
that occurs because the distance to the emitter (and thus the distance
light has to propagate to reach us) is increasing.  In the time
$\delta t_0$ the emitter moves a distance $v\delta t_0$ away from us.
The total proper time we observe, $\delta t_{0,\rm tot}$, is $\delta
t_0$ plus an extra factor describing how long light takes to traverse
this extra distance ($v \delta t_0 /c$),   

\beq 
\delta t_{0,\rm tot} = \delta t_0 (1+v/c). 
\eeq

\noindent
The relationship between proper time at the emitter and proper time at
the observer is thus

\bea
\delta t_{0,\rm tot} &=& \gamma_{\rm SR} \delta t_1 (1+v/c)\\
 &=& \delta t_1 \sqrt{\frac{1+v/c}{1-v/c}} \\ 
 &=& \delta t_1(1+z),
\eea

\noindent
which is identical to the GR time-dilation equation. 

Non-cosmological redshifts (i.e., not due to universal expansion) also
cause a time-dilation effect described by
Eq.~[\ref{eqn:dtfrac}]. However, these additional effects from  
peculiar velocities and gravitational redshifts contribute random
error only, and do not bias the measurements presented here.


\section{Comparison of Spectral and Light-Curve Ages}\label{sect:tcomp}

To test the accuracy of the age determination using SNID, we select
the \sneia\ for which a well-sampled light curve is available around
maximum light. Only the ESSENCE and SNLS \sneia\ in our sample have
associated light curves for which we could determine the date of
maximum brightness ($t_{\rm max}$). To do so we used the MLCS2k2 light-curve
fitting code of \citet{MLCS2k2}, as done by
\cite{Wood-Vasey/etal:2007}. This way we can determine the time
difference (in the {\it observer} frame) between maximum light
($t_{\rm max}$) and the time the spectrum was obtained ($t_{\rm obs}$).
We compare this time interval with the {\it rest-frame} age
determined through cross-correlation with local \snia\ spectral
templates using SNID ($t_{\rm spec}$). We expect a one-to-one
correspondence between

\begin{equation}
t_{\rm LC} = \frac{t_{\rm obs}-t_{\rm max}}{1+z}
\end{equation}

\noindent
and $t_{\rm spec}$. 

\begin{figure}
\epsscale{0.8}
\plotone{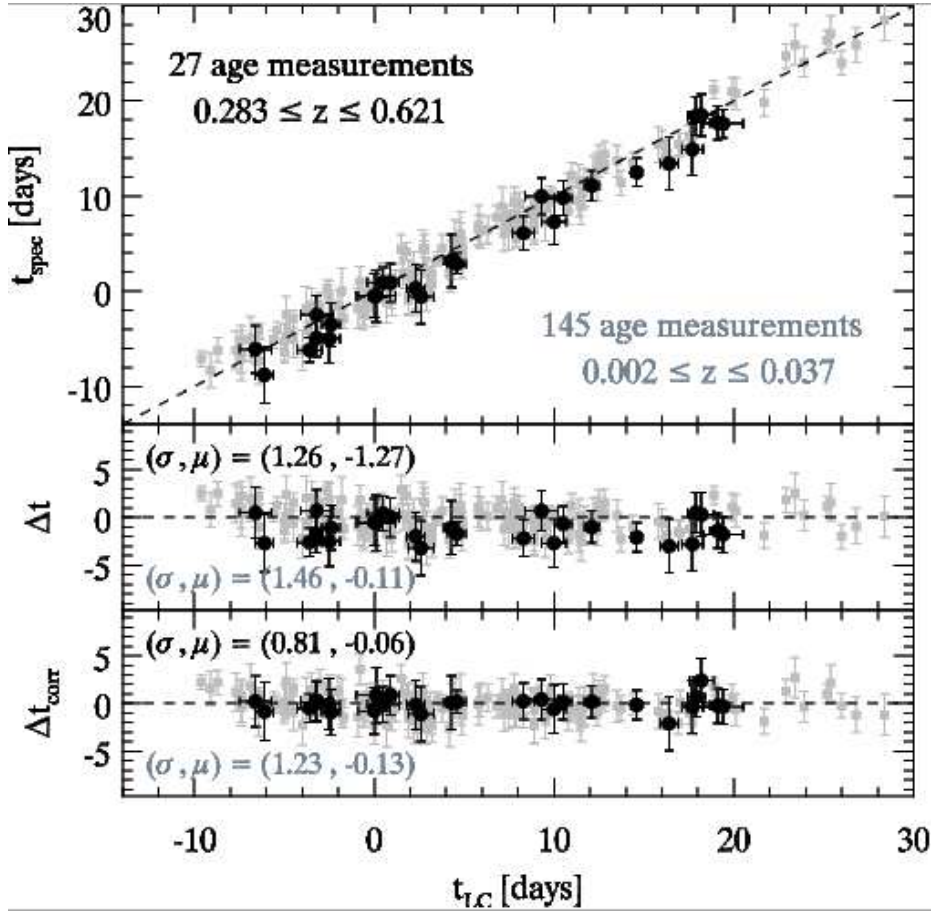}
\caption{\label{fig:tcomp}
Upper panel: comparison of supernova {\it rest-frame} ages (in days
from maximum light) obtained from cross-correlation with spectral
templates ($t_{\rm spec}$) and from fits to the light curve
($t_{\rm LC}$). 145 age measurements for the subsample
of 22 low-redshift \sneia\ are shown in gray. The dashed line
represents the one-to-one correspondence between $t_{\rm LC}$ and
$t_{\rm spec}$.
Middle panel: Age residuals, $\Delta t = t_{\rm spec} - t_{\rm
  LC}$. We also indicate the standard deviation ($\sigma$) and mean
residual ($\mu$).
Lower panel: Same as above, where each point has been corrected for
the mean offset between $t_{\rm spec}$ and $t_{\rm LC}$ for a given
supernova.
}
\end{figure}

The result is shown as black points in Fig.~\ref{fig:tcomp}. While the
agreement is good, there is a mean systematic offset of $-1.6$ d
between $t_{\rm spec}$ and $t_{\rm LC}$, as shown in the middle
panel. If this offset were to affect only a subset of age measurements
for a given supernova, the impact on the aging rate determination
would be severe. To check this, we correct the spectral ages of a
given supernova for the mean difference between $t_{\rm LC}$ and
$t_{\rm spec}$. This ``corrected'' age residual, $\Delta t_{\rm corr}$
is plotted in the lower panel of Fig.~\ref{fig:tcomp}. The mean
residual drops to $-0.1$~d and the scatter decreases slightly. 

Since there are 2 to 4 $t_{\rm spec}$ measurements for a given
supernova, and only one measurement of $t_{\rm max}$, the source of
the discrepancy between the spectral and light-curve ages is most
likely due to the determination of the date of maximum using the
light-curve fitter. Indeed, using a different light-curve fitter
(SALT2; \citealt{SALT2}) yields values for $t_{\rm max}$ that differ from
the MLCS2k2 measurements by more than one day in 9 out of 10 cases,
and by more than two days for three objects (SNe~2003js, 2007tg, and
2007un). These discrepancies are due to a combination of differences
in light-curve fitter algorithms and data quality (light-curve
sampling around maximum light and signal-to-noise ratio of each
light-curve measurement; see \citealt{Miknaitis/etal:2007}).

Therefore, while there is a systematic offset between part of these
different age determinations, this offset affects all measurements in
a similar fashion and has no impact on the determination of the {\it
  rate} of aging. In fact, the main result of this paper (see
Section~\ref{sect:tdil}) is completely independent of $t_{\rm LC}$, and
hence of $t_{\rm max}$. Nonetheless, the comparison between spectral
and light-curve ages confirms the accuracy of age determination using
spectra alone \citep{SNID}. The age measurements for all the
high-redshift \sneia\ in our sample are reported in
Table~\ref{table:tcomp}.

\begin{deluxetable}{lrrrr}
\tablewidth{0pt}
\tablecaption{\label{table:tcomp}Comparison of rest-frame light-curve and
  spectral ages}
\tablehead{
\colhead{SN} &
\colhead{$t_{\rm LC}$} & 
\colhead{$t_{\rm spec}$} & 
\colhead{$\Delta t$} & 
\colhead{$\Delta t_{\rm corr}$} \\
\colhead{(1)} & 
\colhead{(2)} & 
\colhead{(3)} & 
\colhead{(4)} & 
\colhead{(5)}
}
\startdata
2002iz &    0.1 (1.1) &$-$0.5 (2.2) &$-$0.6 (2.4) &   0.8 (1.5) \\
       &   19.4 (1.1) &  17.6 (1.2) &$-$1.8 (1.6) &$-$0.4 (2.3) \\
\hline	 
b027   & $-$2.4 (0.5) &$-$3.5 (1.8) &$-$1.1 (1.9) &$-$0.9 (1.7) \\
       &   17.9 (0.5) &  18.4 (1.6) &   0.5 (1.7) &   0.7 (1.4) \\
\hline	 
2003js & $-$3.2 (0.3) &$-$4.9 (1.6) &$-$1.7 (1.6) &   0.3 (2.0) \\
       &   14.6 (0.3) &  12.5 (1.2) &$-$2.1 (1.2) &$-$0.1 (2.3) \\
\hline
04D2an & $-$3.2 (0.9) &$-$2.5 (1.6) &   0.7 (1.8) &   0.2 (1.4) \\
       &    0.5 (0.9) &   0.9 (1.3) &   0.4 (1.6) &$-$0.1 (1.3) \\
\hline
2006mk & $-$3.6 (0.7) &$-$6.2 (1.0) &$-$2.6 (1.3) &$-$0.5 (2.7) \\
       &    2.6 (0.7) &$-$0.6 (2.2) &$-$3.2 (2.3) &$-$1.1 (3.3) \\
       &   10.0 (0.7) &   7.3 (1.9) &$-$2.7 (2.0) &$-$0.6 (2.8) \\
       &   18.2 (0.7) &  18.5 (1.8) &   0.3 (1.9) &   2.4 (0.9) \\
\hline	 
2006sc &    0.9 (0.5) &   0.9 (1.6) &   0.0 (1.7) &   0.9 (1.0) \\
       &   10.5 (0.5) &   9.8 (1.4) &$-$0.7 (1.5) &   0.2 (1.2) \\
       &   16.4 (0.5) &  13.4 (2.2) &$-$3.0 (2.3) &$-$2.1 (3.2) \\
\hline	 
2006tk & $-$6.1 (0.5) &$-$8.8 (2.4) &$-$2.7 (2.5) &$-$0.8 (2.8) \\
       &    2.3 (0.5) &   0.3 (2.0) &$-$2.0 (2.1) &$-$0.1 (2.2) \\
       &    4.6 (0.5) &   2.9 (0.9) &$-$1.7 (1.0) &   0.2 (1.9) \\
\hline
2007tg & $-$6.6 (0.9) &$-$6.1 (2.0) &   0.5 (2.2) &   0.2 (1.2) \\
       &    0.0 (0.9) &$-$0.5 (1.8) &$-$0.5 (2.0) &$-$0.8 (1.2) \\
       &    9.3 (0.9) &  10.0 (1.5) &   0.7 (1.8) &   0.4 (1.3) \\
\hline
2007tt & $-$2.5 (0.6) &$-$5.0 (2.1) &$-$2.5 (2.2) &$-$0.1 (2.7) \\
       &    8.3 (0.6) &   6.1 (1.4) &$-$2.2 (1.6) &   0.2 (2.5) \\
       &   17.7 (0.6) &  14.9 (2.2) &$-$2.8 (2.2) &$-$0.4 (3.0) \\
\hline
2007un &    4.3 (0.4) &   3.2 (2.2) &$-$1.1 (2.3) &   0.1 (1.4) \\
       &   12.1 (0.4) &  11.1 (1.3) &$-$1.0 (1.3) &   0.2 (1.4) \\
       &   19.1 (0.4) &  17.7 (1.4) &$-$1.4 (1.5) &$-$0.2 (1.7) 
\enddata
\tablenotetext{\ }{{\it Column headings:}
(1) SN name.
(2) SN rest-frame age in days from maximum light, derived from the
  light curve.
(3) SN rest-frame age in days from maximum light,
  derived from the cross-correlation with spectral templates using
  SNID. 
(4) $\Delta t = t_{\rm spec} - t_{\rm LC}$.
(5) $\Delta t$ corrected for the mean offset between $t_{\rm spec}$
  and $t_{\rm LC}$. 
}
\end{deluxetable}

In making the comparison we have implicitly assumed what we are
trying to show, namely a time-dilation factor of $(1+z)$. Accordingly, we
also make the same comparison for our subsample of 22 low-redshift
\sneia\ ($0.002 \le z \le 0.04$). At such low redshifts, the $(1+z)$
correction present in $t_{\rm LC}$ is negligible (the mean correction
is $\sim 0.06$ d). The result is shown as gray points in
Fig.~\ref{fig:tcomp}. The mean residual between $t_{\rm LC}$ and
$t_{\rm spec}$ for this low-redshift sample is close to zero with a
small scatter ($\sigma \approx 1.5$ d), and unlike the
high-redshift sample there is no significant systematic offset between
the two age measurements.

The age measurements presented in Table~\ref{table:tdiff} also
enable us to infer the 
date of maximum light for each supernova using spectra alone
(corresponding to $t_{\rm spec}=0$).
This way we are able to determine the time of
maximum for the \sneia\ in our sample for which a well-sampled
light curve was unavailable (SNe~1996bj, 1997ex, and 2001go; see
Table~\ref{table:tmax}). We can also compare the dates of maximum as
inferred from a fit to the light curve ($t_{\rm max}^{\rm
LC}$) with those determined from the spectra alone ($t_{\rm max}^{\rm
  spec}$). The results are also shown in Table~\ref{table:tmax}. For
four objects (SNe~2003js, 2006mk, 2006tk, and 2007tt) the disagreement is
larger than $1\sigma$, and explains the systematic negative offset
between $t_{\rm spec}$ and $t_{\rm LC}$ seen in Fig.~\ref{fig:tcomp}.

\begin{deluxetable}{lrrr}
\tablewidth{0pt}
\tablecaption{\label{table:tmax}Comparison of dates of maximum light}
\tablehead{
\colhead{SN} &
\colhead{$t_{\rm max}^{\rm LC}$} & 
\colhead{$t_{\rm max}^{\rm spec}$} & 
\colhead{$\Delta t_{\rm max}$} \\
\colhead{(1)} & 
\colhead{(2)} & 
\colhead{(3)} & 
\colhead{(4)}
}
\startdata
1996bj & \nodata        &  372.16 (3.73) &     \nodata     \\
1997ex & \nodata        &  817.16 (1.98) &     \nodata     \\
2001go & \nodata        & 2011.47 (3.12) &     \nodata     \\
2002iz & 2586.83 (1.51) & 2587.71 (3.21) &     0.88 (3.54) \\
b027   & 2593.09 (0.65) & 2594.20 (1.90) &     1.11 (2.01) \\
2003js & 2946.80 (0.47) & 2949.29 (1.67) &     2.49 (1.73) \\
04D2an & 3031.36 (1.50) & 3030.61 (1.82) &  $-$0.75 (2.36) \\
2006mk & 4036.95 (0.96) & 4040.32 (1.02) &     3.37 (1.40) \\
2006sc & 4062.39 (0.64) & 4061.71 (2.77) &  $-$0.68 (2.84) \\
2006tk & 4097.56 (0.66) & 4100.13 (0.98) &     2.57 (1.18) \\
2007tg & 4391.62 (1.34) & 4391.34 (1.58) &  $-$0.28 (2.08) \\
2007tt & 4419.29 (0.83) & 4422.47 (1.83) &     3.18 (2.01) \\
2007un & 4436.12 (0.47) & 4437.15 (3.12) &     1.03 (3.16) 
\enddata
\tablenotetext{\ }{{\it Column headings:}
(1) SN name.
(2) JD $-$ 2,450,000 of maximum light, derived from the light curve.
(3) JD $-$ 2,450,000 of maximum light, derived from the spectra.
(4) $\Delta t_{\rm max} = t_{\rm max}^{\rm spec} - t_{\rm max}^{\rm LC}$.
}
\end{deluxetable}

\end{appendix}


\bibliographystyle{apj}
\bibliography{ms}

\begin{thebibliography}{49}
\expandafter\ifx\csname natexlab\endcsname\relax\def\natexlab#1{#1}\fi

\bibitem[{{Akaike}(1974)}]{AIC}
{Akaike}, H. 1974, IEEE Transactions on Automatic Control, 19, 716

\bibitem[{{Benetti} {et~al.}(2005){Benetti}, {Cappellaro}, {Mazzali},
  {Turatto}, {Altavilla}, {Bufano}, {Elias-Rosa}, {Kotak}, {Pignata}, {Salvo},
  \& {Stanishev}}]{Benetti/etal:2005}
{Benetti}, S., et~al.\ 2005, \apj, 623, 1011

\bibitem[{{Blondin} {et~al.}(2006){Blondin}, {Dessart}, {Leibundgut}, {Branch},
  {H{\"o}flich}, {Tonry}, {Matheson}, {Foley}, {Chornock}, {Filippenko},
  {Sollerman}, {Spyromilio}, {Kirshner}, {Wood-Vasey}, {Clocchiatti},
  {Aguilera}, {Barris}, {Becker}, {Challis}, {Covarrubias}, {Davis},
  {Garnavich}, {Hicken}, {Jha}, {Krisciunas}, {Li}, {Miceli}, {Miknaitis},
  {Pignata}, {Prieto}, {Rest}, {Riess}, {Salvo}, {Schmidt}, {Smith}, {Stubbs},
  \& {Suntzeff}}]{Blondin/etal:2006}
{Blondin}, S., et~al.\ 2006, \aj, 131, 1648

\bibitem[{{Blondin} \& {Tonry}(2007)}]{SNID}
{Blondin}, S. \& {Tonry}, J.~L. 2007, \apj, 666, 1024

\bibitem[{{Conley} {et~al.}(2006){Conley}, {Howell}, {Howes}, {Sullivan},
  {Astier}, {Balam}, {Basa}, {Carlberg}, {Fouchez}, {Guy}, {Hook}, {Neill},
  {Pain}, {Perrett}, {Pritchet}, {Regnault}, {Rich}, {Taillet}, {Aubourg},
  {Bronder}, {Ellis}, {Fabbro}, {Filiol}, {Le Borgne}, {Palanque-Delabrouille},
  {Perlmutter}, \& {Ripoche}}]{Conley/etal:2006b}
{Conley}, A., et~al.\ 2006, \aj, 132, 1707

\bibitem[{{Davis} \& {Lineweaver}(2004)}]{Davis/Lineweaver:2004}
{Davis}, T.~M. \& {Lineweaver}, C.~H. 2004, Publications of the Astronomical
  Society of Australia, 21, 97

\bibitem[{{Davis} {et~al.}(2007){Davis}, {M{\"o}rtsell}, {Sollerman}, {Becker},
  {Blondin}, {Challis}, {Clocchiatti}, {Filippenko}, {Foley}, {Garnavich},
  {Jha}, {Krisciunas}, {Kirshner}, {Leibundgut}, {Li}, {Matheson}, {Miknaitis},
  {Pignata}, {Rest}, {Riess}, {Schmidt}, {Smith}, {Spyromilio}, {Stubbs},
  {Suntzeff}, {Tonry}, {Wood-Vasey}, \& {Zenteno}}]{Davis/etal:2007}
{Davis}, T.~M., et~al.\ 2007, \apj, 666, 716

\bibitem[{{Falco} {et~al.}(1999){Falco}, {Kurtz}, {Geller}, {Huchra}, {Peters},
  {Berlind}, {Mink}, {Tokarz}, \& {Elwell}}]{UZC}
{Falco}, E.~E., et~al.\ 1999, \pasp,  111, 438

\bibitem[{{Foley} {et~al.}(2005){Foley}, {Filippenko}, {Leonard}, {Riess},
  {Nugent}, \& {Perlmutter}}]{Foley/etal:2005}
{Foley}, R.~J., {Filippenko}, A.~V., {Leonard}, D.~C., {Riess}, A.~G.,
  {Nugent}, P., \& {Perlmutter}, S. 2005, \apjl, 626, L11

\bibitem[{{Goldhaber} {et~al.}(2001){Goldhaber}, {Groom}, {Kim}, {Aldering},
  {Astier}, {Conley}, {Deustua}, {Ellis}, {Fabbro}, {Fruchter}, {Goobar},
  {Hook}, {Irwin}, {Kim}, {Knop}, {Lidman}, {McMahon}, {Nugent}, {Pain},
  {Panagia}, {Pennypacker}, {Perlmutter}, {Ruiz-Lapuente}, {Schaefer},
  {Walton}, \& {York}}]{Goldhaber/etal:2001}
{Goldhaber}, G., et~al.\ 2001, \apj, 558, 359

\bibitem[{{Guy} {et~al.}(2007){Guy}, {Astier}, {Baumont}, {Hardin}, {Pain},
  {Regnault}, {Basa}, {Carlberg}, {Conley}, {Fabbro}, {Fouchez}, {Hook},
  {Howell}, {Perrett}, {Pritchet}, {Rich}, {Sullivan}, {Antilogus}, {Aubourg},
  {Bazin}, {Bronder}, {Filiol}, {Palanque-Delabrouille}, {Ripoche}, \&
  {Ruhlmann-Kleider}}]{SALT2}
{Guy}, J., et~al.\ 2007, \aap, 466, 11

\bibitem[{{Hamuy} {et~al.}(1995){Hamuy}, {Phillips}, {Maza}, {Suntzeff},
  {Schommer}, \& {Aviles}}]{Hamuy/etal:1995}
{Hamuy}, M., {Phillips}, M.~M., {Maza}, J., {Suntzeff}, N.~B., {Schommer},
  R.~A., \& {Aviles}, R. 1995, \aj, 109, 1

\bibitem[{{Hamuy} {et~al.}(2003){Hamuy}, {Phillips}, {Suntzeff}, {Maza},
  {Gonz{\'a}lez}, {Roth}, {Krisciunas}, {Morrell}, {Green}, {Persson}, \&
  {McCarthy}}]{Hamuy/etal:2003}
{Hamuy}, M., et~al.\ 2003, \nat, 424, 651

\bibitem[{{Hicken} {et~al.}(2007){Hicken}, {Garnavich}, {Prieto}, {Blondin},
  {DePoy}, {Kirshner}, \& {Parrent}}]{SN2006gz}
{Hicken}, M., {Garnavich}, P.~M., {Prieto}, J.~L., {Blondin}, S., {DePoy},
  D.~L., {Kirshner}, R.~P., \& {Parrent}, J. 2007, \apjl, 669, L17

\bibitem[{{Hook} {et~al.}(2005){Hook}, {Howell}, {Aldering}, {Amanullah},
  {Burns}, {Conley}, {Deustua}, {Ellis}, {Fabbro}, {Fadeyev}, {Folatelli},
  {Garavini}, {Gibbons}, {Goldhaber}, {Goobar}, {Groom}, {Kim}, {Knop},
  {Kowalski}, {Lidman}, {Nobili}, {Nugent}, {Pain}, {Pennypacker},
  {Perlmutter}, {Ruiz-Lapuente}, {Sainton}, {Schaefer}, {Smith}, {Spadafora},
  {Stanishev}, {Thomas}, {Walton}, {Wang}, \& {Wood-Vasey}}]{Hook/etal:2005}
{Hook}, I.~M., et~al.\ 2005, \aj, 130, 2788

\bibitem[{{Howell} {et~al.}(2007){Howell}, {Sullivan}, {Conley}, \&
  {Carlberg}}]{Howell/etal:2007}
{Howell}, D.~A., {Sullivan}, M., {Conley}, A., \& {Carlberg}, R. 2007, \apjl,
  667, L37

\bibitem[{{Howell} {et~al.}(2006){Howell}, {Sullivan}, {Nugent}, {Ellis},
  {Conley}, {Le Borgne}, {Carlberg}, {Guy}, {Balam}, {Basa}, {Fouchez}, {Hook},
  {Hsiao}, {Neill}, {Pain}, {Perrett}, \& {Pritchet}}]{03D3bb}
{Howell}, D.~A., et~al.\ 2006, \nat, 443, 308

\bibitem[{{Howell} {et~al.}(2005){Howell}, {Sullivan}, {Perrett}, {Bronder},
  {Hook}, {Astier}, {Aubourg}, {Balam}, {Basa}, {Carlberg}, {Fabbro},
  {Fouchez}, {Guy}, {Lafoux}, {Neill}, {Pain}, {Palanque-Delabrouille},
  {Pritchet}, {Regnault}, {Rich}, {Taillet}, {Knop}, {McMahon}, {Perlmutter},
  \& {Walton}}]{Howell/etal:2005}
{Howell}, D.~A., et~al.\ 2005, \apj, 634, 1190

\bibitem[{{Jha} {et~al.}(2007){Jha}, {Riess}, \& {Kirshner}}]{MLCS2k2}
{Jha}, S., {Riess}, A.~G., \& {Kirshner}, R.~P. 2007, \apj, 659, 122

\bibitem[{{Leibundgut} {et~al.}(1996){Leibundgut}, {Schommer}, {Phillips},
  {Riess}, {Schmidt}, {Spyromilio}, {Walsh}, {Suntzeff}, {Hamuy}, {Maza},
  {Kirshner}, {Challis}, {Garnavich}, {Smith}, {Dressler}, \&
  {Ciardullo}}]{SN1995K}
{Leibundgut}, B., et~al.\ 1996, \apjl, 466, L21

\bibitem[{{Li} {et~al.}(2003){Li}, {Filippenko}, {Chornock}, {Berger},
  {Berlind}, {Calkins}, {Challis}, {Fassnacht}, {Jha}, {Kirshner}, {Matheson},
  {Sargent}, {Simcoe}, {Smith}, \& {Squires}}]{Li/etal:2003}
{Li}, W., et~al.\ 2003, \pasp, 115, 453

\bibitem[{{Li} {et~al.}(2001{\natexlab{a}}){Li}, {Filippenko}, {Gates},
  {Chornock}, {Gal-Yam}, {Ofek}, {Leonard}, {Modjaz}, {Rich}, {Riess}, \&
  {Treffers}}]{Li/etal:2001b}
{Li}, W., et~al.\ 2001{\natexlab{a}}, \pasp, 113, 1178

\bibitem[{{Li} {et~al.}(2001{\natexlab{b}}){Li}, {Filippenko}, {Treffers},
  {Riess}, {Hu}, \& {Qiu}}]{Li/etal:2001a}
{Li}, W., {Filippenko}, A.~V., {Treffers}, R.~R., {Riess}, A.~G., {Hu}, J., \&
  {Qiu}, Y. 2001{\natexlab{b}}, \apj, 546, 734

\bibitem[{{Liddle}(2004)}]{Liddle:2004}
{Liddle}, A.~R. 2004, \mnras, 351, L49

\bibitem[{{Lidman} {et~al.}(2005){Lidman}, {Howell}, {Folatelli}, {Garavini},
  {Nobili}, {Aldering}, {Amanullah}, {Antilogus}, {Astier}, {Blanc}, {Burns},
  {Conley}, {Deustua}, {Doi}, {Ellis}, {Fabbro}, {Fadeyev}, {Gibbons},
  {Goldhaber}, {Goobar}, {Groom}, {Hook}, {Kashikawa}, {Kim}, {Knop}, {Lee},
  {Mendez}, {Morokuma}, {Motohara}, {Nugent}, {Pain}, {Perlmutter}, {Prasad},
  {Quimby}, {Raux}, {Regnault}, {Ruiz-Lapuente}, {Sainton}, {Schaefer},
  {Schahmaneche}, {Smith}, {Spadafora}, {Stanishev}, {Walton}, {Wang},
  {Wood-Vasey}, {Yasuda}, \& {The Supernova Cosmology
  Project}}]{Lidman/etal:2005}
{Lidman}, C., et~al.\ 2005, \aap, 430, 843

\bibitem[{{Matheson} {et~al.}(2005){Matheson}, {Blondin}, {Foley}, {Chornock},
  {Filippenko}, {Leibundgut}, {Smith}, {Sollerman}, {Spyromilio}, {Kirshner},
  {Clocchiatti}, {Aguilera}, {Barris}, {Becker}, {Challis}, {Covarrubias},
  {Garnavich}, {Hicken}, {Jha}, {Krisciunas}, {Li}, {Miceli}, {Miknaitis},
  {Prieto}, {Rest}, {Riess}, {Salvo}, {Schmidt}, {Stubbs}, {Suntzeff}, \&
  {Tonry}}]{Matheson/etal:2005}
{Matheson}, T., et~al.\ 2005, \aj, 129,  2352

\bibitem[{{Matheson} {et~al.}(2008){Matheson}, {Kirshner}, {Challis}, {Jha},
  {Garnavich}, {Berlind}, {Calkins}, {Blondin}, {Balog}, {Bragg}, {Caldwell},
  K., {Falco}, {Graves}, \& {Huchra}}]{Matheson/etal:2008}
{Matheson}, T., et~al.\ 2008, accepted for publication in \aj

\bibitem[{{Mazzali} {et~al.}(2008){Mazzali}, {Sauer}, {Pastorello}, {Benetti},
  \& {Hillebrandt}}]{Mazzali/etal:2008}
{Mazzali}, P.~A., {Sauer}, D.~N., {Pastorello}, A., {Benetti}, S., \&
  {Hillebrandt}, W. 2008, ArXiv e-prints, 803

\bibitem[{{Miknaitis} {et~al.}(2007){Miknaitis}, {Pignata}, {Rest},
  {Wood-Vasey}, {Blondin}, {Challis}, {Smith}, {Stubbs}, {Suntzeff}, {Foley},
  {Matheson}, {Tonry}, {Aguilera}, {Blackman}, {Becker}, {Clocchiatti},
  {Covarrubias}, {Davis}, {Filippenko}, {Garg}, {Garnavich}, {Hicken}, {Jha},
  {Krisciunas}, {Kirshner}, {Leibundgut}, {Li}, {Miceli}, {Narayan}, {Prieto},
  {Riess}, {Salvo}, {Schmidt}, {Sollerman}, {Spyromilio}, \&
  {Zenteno}}]{Miknaitis/etal:2007}
{Miknaitis}, G., et~al.\ 2007,  \apj, 666, 674

\bibitem[{{Nomoto} {et~al.}(1984){Nomoto}, {Thielemann}, \& {Yokoi}}]{W7}
{Nomoto}, K., {Thielemann}, F.-K., \& {Yokoi}, K. 1984, \apj, 286, 644

\bibitem[{{Nugent} {et~al.}(1995){Nugent}, {Phillips}, {Baron}, {Branch}, \&
  {Hauschildt}}]{Nugent/etal:1995}
{Nugent}, P., {Phillips}, M., {Baron}, E., {Branch}, D., \& {Hauschildt}, P.
  1995, \apjl, 455, L147+

\bibitem[{{Phillips}(1993)}]{Phillips:1993}
{Phillips}, M.~M. 1993, \apjl, 413, L105

\bibitem[{{Phillips} {et~al.}(1999){Phillips}, {Lira}, {Suntzeff}, {Schommer},
  {Hamuy}, \& {Maza}}]{Phillips/etal:1999}
{Phillips}, M.~M., {Lira}, P., {Suntzeff}, N.~B., {Schommer}, R.~A., {Hamuy},
  M., \& {Maza}, J. 1999, \aj, 118, 1766

\bibitem[{{Pinto} \& {Eastman}(2000)}]{Pinto/Eastman:2000b}
{Pinto}, P.~A. \& {Eastman}, R.~G. 2000, \apj, 530, 757

\bibitem[{{Riess} {et~al.}(1997){Riess}, {Filippenko}, {Leonard}, {Schmidt},
  {Suntzeff}, {Phillips}, {Schommer}, {Clocchiatti}, {Kirshner}, {Garnavich},
  {Challis}, {Leibundgut}, {Spyromilio}, \& {Smith}}]{Riess/etal:1997a}
{Riess}, A.~G., et~al.\ 1997, \aj, 114, 722

\bibitem[{{Riess} {et~al.}(1999){Riess}, {Filippenko}, {Li}, {Treffers},
  {Schmidt}, {Qiu}, {Hu}, {Armstrong}, {Faranda}, {Thouvenot}, \&
  {Buil}}]{Riess/etal:1999c}
{Riess}, A.~G., et~al.\ 1999, \aj, 118, 2675

\bibitem[{{Riess} {et~al.}(1995){Riess}, {Press}, \&
  {Kirshner}}]{Riess/Press/Kirshner:1995}
{Riess}, A.~G., {Press}, W.~H., \& {Kirshner}, R.~P. 1995, \apjl, 438, L17

\bibitem[{{Robertson}(1935)}]{Robertson:1935}
{Robertson}, H.~P. 1935, \apj, 82, 284

\bibitem[{{Robertson}(1936{\natexlab{a}})}]{Robertson:1936a}
---. 1936{\natexlab{a}}, \apj, 83, 187

\bibitem[{{Robertson}(1936{\natexlab{b}})}]{Robertson:1936b}
---. 1936{\natexlab{b}}, \apj, 83, 257

\bibitem[{{Rust}(1974)}]{Rust:1974}
{Rust}, B.~W. 1974, PhD thesis, Oak Ridge National Lab., TN.

\bibitem[{{Stehle} {et~al.}(2005){Stehle}, {Mazzali}, {Benetti}, \&
  {Hillebrandt}}]{Stehle/etal:2005}
{Stehle}, M., {Mazzali}, P.~A., {Benetti}, S., \& {Hillebrandt}, W. 2005,
  \mnras, 360, 1231

\bibitem[{{Stritzinger}(2005)}]{Stritzinger:2005}
{Stritzinger}, M. 2005, PhD thesis, Technische Universit{\"a}t M{\"u}nchen,

\bibitem[{{Tonry} \& {Davis}(1979)}]{Tonry/Davis:1979}
{Tonry}, J. \& {Davis}, M. 1979, \aj, 84, 1511

\bibitem[{{Walker}(1936)}]{Walker:1936}
{Walker}, A.~G. 1936, in Proceedings of the London Mathematical Society,
  90--127

\bibitem[{{Weinberg}(1972)}]{Weinberg:1972}
{Weinberg}, S. 1972, {Gravitation and Cosmology: Principles and Applications of
  the General Theory of Relativity} (Gravitation and Cosmology: Principles and
  Applications of the General Theory of Relativity, by Steven Weinberg,
  pp.~688.~ISBN 0-471-92567-5.~Wiley-VCH , July 1972.)

\bibitem[{{Wilson}(1939)}]{Wilson:1939}
{Wilson}, O.~C. 1939, \apj, 90, 634

\bibitem[{{Wood-Vasey} {et~al.}(2007){Wood-Vasey}, {Miknaitis}, {Stubbs},
  {Jha}, {Riess}, {Garnavich}, {Kirshner}, {Aguilera}, {Becker}, {Blackman},
  {Blondin}, {Challis}, {Clocchiatti}, {Conley}, {Covarrubias}, {Davis},
  {Filippenko}, {Foley}, {Garg}, {Hicken}, {Krisciunas}, {Leibundgut}, {Li},
  {Matheson}, {Miceli}, {Narayan}, {Pignata}, {Prieto}, {Rest}, {Salvo},
  {Schmidt}, {Smith}, {Sollerman}, {Spyromilio}, {Tonry}, {Suntzeff}, \&
  {Zenteno}}]{Wood-Vasey/etal:2007}
{Wood-Vasey}, W.~M., et~al.\ 2007, \apj, 666, 694

\bibitem[{{Zwicky}(1929)}]{Zwicky:1929}
{Zwicky}, F. 1929, Proceedings of the National Academy of Science, 15, 773


\end{thebibliography}

\end{document}